\documentclass[a4paper, 11pt]{article}

\usepackage{amsmath, amsthm}
\usepackage{amsfonts, latexsym}
\usepackage{amssymb}
\usepackage[dvips]{graphicx}
\usepackage{psfrag}
\usepackage{hyperref}
\usepackage[T1]{fontenc}

\newcommand{\vect}[1]{\textbf{#1}}
\def\i{\textnormal{i}}
\def\e{\textnormal{e}}
\def\d{\textnormal{d}}

\begin{document}

\title{\bf Fractional Fourier Transform\\ and Geometric Quantization}

\author{
  Witold Chmielowiec\footnote{e-mail: wchmiel@cft.edu.pl} 
  and Jerzy Kijowski\footnote{e-mail: kijowski@cft.edu.pl}
  \\
  Center for Theoretical Physics, Polish Academy of Sciences\\
  Al. Lotnik\'ow 32/46, 02-668 Warsaw, Poland
  }

\date{ }

\maketitle

\begin{abstract}

Generalized Fourier transformation between the position and the
momentum representation of a quantum state is constructed in a
coordinate independent way. The only ingredient of this
construction is the symplectic (canonical) geometry of the
phase-space: no linear structure is necessary. It is shown that
the ``fractional Fourier transform''
provides a simple example of this construction. As an application
of this techniques we show that for any linear Hamiltonian system,
its quantum dynamics can be obtained {\em exactly} as the lift of
the corresponding classical dynamics by means of the above
transformation. Moreover, it can be deduced from the {\em free
quantum evolution}. This way new, unknown symmetries of the
Schr\"odinger equation can be constructed. It is also argued that the above
construction defines in a natural way a connection in the bundle of quantum
states, with the base space describing all their possible representations. The
non-flatness of this connection would be responsible for the non-existence of 
a quantum representation of the complete algebra of classical observables.

\end{abstract}



\section{Introduction}
\label{sec:Introduction}

Correct mathematical description of a specific quantum system was
in many cases obtained {\em via} an appropriate ``quantization
procedure'' from the corresponding ``classical theory''. The first
example of this type is the Heisenberg approach to quantum
mechanics. Most of the field-theoretical models, like quantum
electrodynamics, have also been constructed this way.

Geometric quantization (see e.~g.~\cite{Sou}) was an attempt to
formalize the above analogy between classical and quantum systems
and to provide a tool to construct uniquely quantum theory once
its classical counterpart is known. In particular, it has been
noticed that some mathematical structures used in mechanics and
classical field theory on one side and in quantum mechanics and
quantum field theory on the other are very similar. It was obvious
from the very beginning that this analogy cannot go too far
because quantum physics {\em cannot be reduced} to classical
physics. Nevertheless, such a unifying point of view has lead to
important mathematical results in theory of group representations,
theory of analytic functions, differential geometry and other
branches of mathematics (see e.~g.~\cite{Kos}).

Seen from the physical context, quantum mechanics (both the Schr\"o\-din\-ger
and the Heisenberg version) has, {\em a priori} very little to do with the
symplectic structure of the underlying classical phase space. Indeed, it {\em
is not} invariant with respect to (non-linear) canonical transformations!

{\em A priori} it is even non-invariant with respect to non-linear point
transformations! However, this invariance may be easily restored if we use
metric structure $g=g_{kl}\d x^k \d x^l$ of the configuration space $Q$,
define the Hilbert space of pure states as
${\cal H}:=L^2(Q, \sqrt{\det g_{kl}}\,\cdot \mbox{\rm d}^n x)$ and take the
Laplace-Beltrami operator (with minus sign) as the kinetic energy.

This is a relatively nice framework (called ``covariant quantum mechanics''),
which was recently thoroughly analyzed e.g. by M.~Modugno (cf.
\cite{Modugno}). In particular, no assumptions concerning the  topology of the
configuration space $Q$ are necessary here. However, there are severe
restrictions for the applicability of this approach: the time must be
absolute, only non-relativistic Hamiltonians (i.e. "kinetic plus potential
energy") are allowed, no momentum representation is available etc.

A deep analysis of quantum mechanics and relativistic quantum
field theory, first performed by J.~M.~Souriau (cf.~\cite{Sou})
and, independently, by W.~Tulczyjew (cf.~\cite{Tulczyjew}), lead
to the formulation of ``geometric quantization theory'', based on
the phase space and its symplectic structure.

A popular approach to ``geometric quantization'' consists in
defining the quantum dynamics in terms of the ``reproducing
kernels'' (see \cite{Gaw, Kij3} and the monographs \cite{Wood} or
\cite{Snia}), which are carried by
the geometric structure of the phase space of the system with
finite number of degrees of freedom (The corresponding structures
arising in field theory was analyzed in \cite{Kij2} and
\cite{Kij1}).

Unfortunately, the complete symplectic structure cannot be represented on the
quantum level, even if more and more sophisticated mathematical tools are
introduced. In particular, the classical observable algebra (equipped with the
Poisson bracket) has no appropriate {\em irreducible} representation in the
algebra of operators acting in the Hilbert space. Here, ``appropriate'' means
that it reduces to the standard quantum mechanics when restricted to the
(finite or infinite-dimensional) Heisenberg algebra. Different
functional-analytic frameworks can be chosen in order to convert the above
``meta-mathematical'' statement into a precise theorem, but non of them (e.g.:
bounded or unbounded, continuous, smooth or only measurable observables) leads
to a satisfactory representation. The notion of a {\em prequantization},
introduced by J.~M.~Souriau, even if mathematically beautiful, does not help
much, because it leads to the representation which is highly {\em reducible}
and, therefore, cannot be used when calculating e.g. atomic optical spectra.

Physicists, chemists and quantum opticians, who try to model physical
properties of complicated multi-molecular systems {\em via} Schr\"odinger
equation, often use specific ``quantization rules'', formulated in terms of
specific ``orderings'' imposed on products of operators (e.g.: normal,
anti-normal, Weyl etc., cf. \cite{Penson} and references therein). They
observe that the calculated spectra depend upon the ordering chosen. This fact
may be considered as the ``practical proof'' that the entire classical
observable algebra has no quantum representation.

There are deep mathematical results due to geometric quantization theory
(e.g.: representation theory, ``reproducing kernels'', metaplectic structure,
the Maslov index etc.). These results are based on highly sophisticated
mathematical tools. In spite of that, they are not very useful for
applications. As a consequence, they remain virtually unknown to physicists.
On the other hand, quantum physicists often discover some elements of this
symplectic {\em Atlantis} but, in most cases, they are not fully aware of the
consequences of these discoveries. This was recently the case of the
``fractional Fourier transform'', an old mathematical structure rediscovered
in quantum optics.

The goal of our paper is to present the basic structures which are
necessary to formulate quantum mechanics in a simplest language
and to analyze the symplectic invariance of the theory. To make
our presentation as simple as possible we limit ourselves to the
topologically trivial case (i.e.~when the physical phase space is
topologically equal to $\mathbb{R}^{2n}$). This is the case of
most physical applications. Moreover, we use only those
representations of the quantum Hilbert space which correspond to
the so called ``real polarizations''. This excludes some
interesting issues like e.g. Bargmann representation, but allows
us to simplify considerably the mathematical framework which does
not go beyond the geometric interpretation of what the physicists
know from the very beginning of quantum mechanics and may be found
in standard textbooks.

Even if mathematically not sophisticated, our approach describes
all the essential features of quantum mechanics. In particular, we
prove that in case of linear dynamics, geometric
quantization cannot fail: the {\em entire information} about
quantum dynamics can be retrieved from its classical counterpart.
We show that the correct evolution kernels can be obtained from
the classical dynamics {\em via} a universal formula which is
nothing but a~properly geometrized Fourier transformation,
superposed with (again: properly geometrized) Galilei
transformation. In fact, these two transformations are the only ``building 
blocks'' of our approach. We show, that the complete description of quantum 
mechanics may be obtained if we use them in a correct way.

In particular case of a harmonic oscillator, the
``fractional Fourier transform'' is obtained as a~specific
example. This way we prove that dynamics of various quantum
systems, which look apparently very different (like e.~g.: free
motion, harmonic oscillator, motion in a constant electric or
magnetic field) provide specific examples of a single, universal
formula. Moreover, the classical isomorphism relating any two
cases of classical linear dynamics on the quantum level is represented by
a~{\em local} (with respect to space and time)
isomorphism between the corresponding Hilbert spaces. In
particular, unexpected symmetries of the Schr\"odinger equation
are obtained (some of them were known already long time ago, see
e.~g.~\cite{KijJak} and references herein). Because both the
Fourier transformation and the Galilei transformation describe
statics (change of the representation and change of the reference
frame) we conclude that the dynamics of linear systems is entirely
implied by their static properties.

In the last part of the paper we show how a generic, non-linear
classical evolution can be lifted to the quantum evolution {\em
via} a natural connection in the bundle of quantum states. The
connection is, however, non-flat  and this is why the entire
canonical structure of the phase space cannot be represented on
the quantum level.

\section{Fractional Fourier transform}
\label{sec:FFThod}

The fractional Fourier transform (FrFT) is known e.g.~from Namias
paper \cite{Nam}. It gives an important tool in classical optics \cite{Loh1,
Loh2, Oza1, Oza2}, quantum optics \cite{Ali, Chou, Lu1, Lu2, Tas, Wal}
and signal processing \cite{Alm}. But the idea of such
an integral transformation appeared much earlier in mathematical
literature, see e.g.~\cite{Kob}. The 1-dimensional FrFT is given
by the following formula (the coefficients have been chosen in
a way which is suitable for purposes of quantum mechanics):
\begin{equation}\label{eq:FrFT_df}
    (\mathcal{F}_\gamma f)(x') =
    \int K(\gamma,x,x') f(x)  \, \d x,
\end{equation}
where the kernel $K(\gamma,x,x')$ is given by
\begin{equation}\label{eq:FrFT}
    K(\gamma,x,x') = \frac{\e^{\i\frac \gamma 2}}{\sqrt{\i\sin\gamma}}
    \  \e^{\i\pi
    \left((x^2+x'^2)\cot\gamma-\frac{2x'x}{\sin\gamma}\right)} \ .
\end{equation}
Here, $f$ is a complex-valued function (in applications $f$ can
describe a quan\-tum-mechanical wave-function, or a fully
coherent, quasimonochromatic, classical electromagnetic wave). The
constant $\gamma$ is a real number \cite{Nam, McB}. The formula
\eqref{eq:FrFT} is, {\em a priori} meaningless for $\gamma=0$ but
its limit for $\gamma \rightarrow 0$ does exist and is equal to
the Dirac distribution $\delta(x - x')$. Hence, the corresponding
limit of the transformation \eqref{eq:FrFT_df} is equal to
identity: $\mathcal{F}_0 f = f$.

The transformation is called ``fractional'', because it provides
an interpolation between the identity operator $\mathcal{F}_0$ and
the ordinary Fourier transform which we obtain for $\gamma=\frac
\pi 2$. Indeed, operators $\mathcal{F}_\gamma$ depend continuously
upon the parameter $\gamma$ and satisfy the group property (see
\cite{Nam,McB}):
\begin{equation}\label{eq:add}
  \mathcal{F}_{\mu+\nu}=\mathcal{F}_\mu \mathcal{F}_\nu.
\end{equation}

Observe that the formula for the quantum-mechanical propagator of
the harmonic oscillator with frequency $\omega$ and mass $m$:
\begin{equation}\label{eq:prop}
    G(t,x,x') = \sqrt{\frac{m\omega}{2\pi\i\hbar\sin\omega t}}
    \  \e^{\frac{\i m\omega}{2\hbar}
    \left((x^2+x'^2)\cot\omega t-\frac{2x'x}{\sin\omega t}\right)},
\end{equation}
reduces (up to a constant phase factor) to  \eqref{eq:FrFT} if we
choose $\gamma = \omega t$ and re-scale appropriately coordinates
$x$, $x'$. This observation is already known from the Namias' work
\cite{Nam}. In the present paper we show that this transformation
is a specific example of a Generalized Fourier Transformation
which will be defined in a purely geometric, coordinate-invariant
way.

Before we present this construction in subsequent Sections, we are
going to show that formula \eqref{eq:prop} can be simply
understood as a superposition of the following two standard
operations: 1) the conventional Fourier transformation between the
position and the momentum representations and 2) the Galilei
transformation changing the phase of the wave function as a
consequence of the change of a reference frame.

To prove the above statement let us consider the classical
dynamics of the harmonic oscillator\footnote{We use here the
``Heisenberg picture'': points of the phase space do not move
during the evolution and represent entire {\em histories} of the
system. Evolution applies to {\em observables}. Hence,
$(x(0),p(0))$ and $(x(t), p(t))$ have to be understood as two
different coordinate systems in the same phase space. Souriau
calls this phase space ``espace des mouvements''.}:
\begin{equation}\label{eq:osc_solve-a}
  \begin{split}
      x(t) &= x(0)\, \cos\omega t + p(0)\, \frac{1}{m\omega} \sin\omega t
     \ ,\\
      p(t) &= - x(0)\, m\omega  \sin\omega t + p(0)\, \cos\omega t
      \ .
  \end{split}
\end{equation}
Denote $x := x(0)$, $p := p(0)$, $x' := x(t)$ and $p' := p(t)$.
Consider first the particular case $\omega t = \frac \pi 2$. We
have:
\begin{equation}\label{particular}
       x' = \frac p{m \omega} \ .
\end{equation}
Hence, $p=m \omega x' $ is the momentum canonically conjugate to
$x$. Therefore, transition between the $x(0)$-representation and
$x(t)$-representation of the quantum state must be given in terms
of the transition between the position and the momentum
representation. Indeed, formula \eqref{eq:prop} reduces to
\begin{equation}\label{eq:prop-a}
    G(t,x,\frac p{m \omega}) = \sqrt{\frac{m\omega}{2\pi\i\hbar}}
    \  \e^{-\frac{\i px}{\hbar} } =
    \sqrt{- \i m \omega} \ \sqrt{\frac{1}{2\pi\hbar}}
    \  \e^{-\frac{\i px}{\hbar} }
    \ ,
\end{equation}
which, essentially, is the Fourier kernel defining the transition
to the momentum representation. However, we have an extra
coefficient ``$\sqrt{- \i m\omega}$'' on the right hand side. Its
constant phase factor ``$\sqrt{- \i}$'' is due to the convention
used and has no physical meaning. But its modulus is necessary
because the wave function is not a scalar object but a {\em
half-density}\footnote{In traditional courses of differential geometry, like 
e.~g.~\cite{Schouten}, half-densities were called {\em scalar densities of 
weight} $\tfrac 12$. For a modern definition and examples see also 
\cite{Abraham-Marsden}. The value of a half-density at a point of an 
$n$-dimensional manifold, is a positively homogeneous (of degree $\tfrac 12$) 
function on the space of $n$-vectors attached at this point. For some purposes 
(e.~g.~{\em metaplectic group}) people distinguish between ``non-oriented'' 
and ``oriented'' objects, the latter being often called ``half-forms''. Here, 
we use the simplest, non-oriented objects.}. Without going too far into 
mathematical subtleties,
which will became obvious in the next Section, the above statement
means that the square of the modulus of a wave function is a
density. Hence, the coefficient ``$\sqrt{ m\omega}$'' is necessary
because its square ``$m\omega$'' represents the change of the
volume due to the reparametrization $p \mapsto x' =\frac p{m
\omega}$ of the momentum space.

Now, consider an arbitrary value of the time variable $t$. The same formula
(namely: $x' = x\, \cos\omega t + p\, \frac{1}{m\omega} \sin\omega t$) can be
rewritten as:
\begin{equation}\label{p-tylda}
  p + x \cdot m \omega  \cot \omega t = \frac {m\omega}{\sin\omega t} \ x'
  =: \tilde{p} \ .
\end{equation}
We conclude that the quantity $\tilde{p}$ may be taken as a
momentum canonically conjugate to $x$. The argument used above
explains the multiplicative factor $\sqrt{\frac
{m\omega}{\sin\omega t}}$ in formula \eqref{eq:prop} and the last
term in the exponent. But, there is an additional phase factor,
namely $\exp \left(\frac{\i m\omega}{2\hbar} (x^2+x'^2)\cot\omega
t\right)$. We are going to show that it is a consequence of the
Galilei transformation corresponding to formula \eqref{p-tylda}.

Indeed, formula \eqref{p-tylda} is a particular example of
a canonical transformation between the old canonical variables $(x,p)
\mapsto (x,\tilde{p})$, where the new momentum is given by:
\begin{equation}\label{GAL}
  \tilde{p} = p + F(x) \ .
\end{equation}
Such a ``momentum translation''  arises e.g.~when performing a Galilei
transformation:
\[
  \tilde{x} := x - t \cdot V \ ,
\]
where $V$ denotes the velocity of the new reference frame. Consequently, we
have $\dot{\tilde{x}} = \dot{x} - V$ and, therefore,
\[
   \tilde{p} = m \dot{\tilde{x}} = m (\dot{x} - V) = p - mV \ ,
\]
whereas $\tilde{x} = x$ at $t=0$.

Transformation \eqref{GAL} is called a {\em generalized Galilei
transformation}, the name {\em proper Galilei transformation} being reserved
for the case when the function $F(x)$ is constant.

In a generic, multidimensional case, transformation $(x^i,p_i) \mapsto
(x^i,\tilde{p}_i)$, with
\begin{equation}\label{GAL-1}
  \tilde{p}_i = p_i + F_i(x) \ .
\end{equation}
is canonical if and only if the differential 1-form $\alpha := F_i \mbox{\rm
d} x^i$ is closed. Due to topological triviality of the configuration space
this is equivalent to the fact that  $\alpha$ must be exact, i.e.~we have:
\begin{equation}\label{F-grad}
    F_i(x) = \frac{\partial}{\partial x^i} S(x) \ .
\end{equation}

The quantum version of the generalized Galilei transformation \eqref{GAL-1} is
obvious. It consists in multiplying the wave
function by the phase factor $\exp \left( \frac \i\hbar S(x)
\right)$:
\begin{equation}\label{GAL-Quantum}
    \tilde{\psi}(x) := \psi(x) \cdot \exp \left( \frac \i\hbar S(x)
\right) \ ,
\end{equation}
which, together with the Schr\"odinger representation of the momenta
\begin{equation}\label{rep-pi}
      p_i = \frac \hbar\i \frac{\partial}{\partial x^i} \ ,
\end{equation}
reproduces, indeed, formula  \eqref{GAL-1}. We stress that the phase $S$ is
implied by \eqref{F-grad} up to an additive constant only. This agrees with
the fact that the global phase of the wave function has no physical
significance.

In particular case of the transformation \eqref{p-tylda}, we have $F(x) = x
\cdot m \omega  \cot \omega t$ and, therefore:
\begin{equation}\label{faza}
    S(x) = \frac {m \omega}2 \ x^2 \ \cot \omega t \ ,
\end{equation}
which explains the phase factor $\exp \left(\frac{\i m\omega}{2\hbar}
x^2\cot\omega t\right)$ in formula \eqref{eq:prop}.

To explain the remaining phase factor, namely:
$\exp \left(\frac{\i m\omega}{2\hbar} x'^2\cot\omega
t\right)$, let us first summarize the sequence of operations which have to be
applied to the wave function $\psi(x)$ in order to reproduce the
transformation defined by the integral kernel \eqref{eq:prop}.

\begin{enumerate}
  \item Wave function $\psi(x)$ represents the quantum state with respect to
  the Heisenberg algebra generated by observables $(x,p)$. Its representation
  $\tilde{\psi}(x)$ with respect to $(x, \tilde{p})$, where $\tilde{p}= p +
  \frac{\mbox{\rm d}}{\mbox{\rm d} x} S(x)$, is obtained {\em via} the
  generalized Galilei transformation \eqref{GAL-Quantum}:
  \begin{equation}
    \tilde{\psi}(x) := \psi(x) \cdot
    \exp \left(\frac{\i m\omega}{2\hbar} x^2\cot\omega t\right) \;.
  \end{equation}

  \item We pass to the momentum representation using the ordinary Fourier
  transformation $\hat{\tilde{\psi}}(\tilde{p})$ of the function
  $\tilde{\psi}(x)$. This way we exchange the role of $x$ and $\tilde{p}$,
  which corresponds to the canonical transformation $(x, \tilde{p})
  \mapsto  (\tilde{p}, -x)$. The momentum canonically conjugate to
  $\tilde{p}$, namely   $-x$, is represented now by the operator $\frac \hbar
  \i \frac{\partial}{\partial \tilde{p}}\,$ acting on the wave
  function~$\hat{\tilde{\psi}}(\tilde{p})$.

  \item Next step consists in using Ansatz \eqref{p-tylda}, i.e.: $\tilde{p}
  := \frac {m\omega}{\sin\omega t} \ x' $. This means that we implement the
  canonical transformation:
   \[
      \left(\frac {m\omega}{\sin\omega t} \; x', -x \right) \mapsto
      \left(x', - \frac {m\omega}{\sin\omega t}\; x\right) \ .
   \]
   For this purpose only the density factor $\sqrt{\frac {m\omega}
   {\sin\omega t}}$ is necessary and we obtain the new wave function:
   \begin{equation}\label{compensate}
     \phi(x') := \sqrt{\frac {m\omega}{\sin\omega t}} \cdot
     \hat{\tilde{\psi}}\left(\frac {m\omega}{\sin\omega t} \ x'\right)
   \end{equation}

   \item Finally, we want to replace the ``fictitious'' momentum
   $q:=- \frac {m\omega}{\sin\omega t}\;x$ by the ``true'' momentum $p'$,
   canonically conjugate to $x'$. For this purpose we use again formulae
   \eqref{eq:osc_solve-a} and \eqref{p-tylda}:
   \begin{align*}
       p' &= -x\cdot m\omega \sin \omega t + p\cdot \cos \omega t
     = - x \cdot \frac{m\omega}{\sin\omega t} + x' \cdot m \omega
     \cot \omega t
     \\
     &= q +  x' \cdot m \omega \cot \omega t \;.
   \end{align*}
  We see that, again, a Galilei transformation is necessary, with the phase
  factor $\exp \left( \frac i\hbar S(x') \right)$ defined by equation
  \[
    p' - q = x' \cdot m \omega \cot \omega t =
    \frac{\mbox{\rm d}}{\mbox{\rm d} x'} S(x') \ ,
  \]
  and, whence, given by formula \eqref{faza}. This way we obtain the final
  wave function
  \begin{equation}\label{GAL-Quantum-3}
    \psi'(x') := \phi(x') \cdot \exp \left(\frac{\i m\omega}{2\hbar}x'^2
    \cot\omega t \right) \ .
  \end{equation}
  We conclude that the entire missing phase factor in formula \eqref{eq:prop}
  comes from the above Galilei transformation.
\end{enumerate}

The above procedure shows that the ``fractional Fourier transform'' kernel
\eqref{eq:prop} is nothing but the {\em ordinary} Fourier kernel (step 2.),
appropriately superposed with two Galilei transformations (steps 1. and 4.)
and one obvious transformation coming from rescaling of the corresponding
configuration space (step 3.). These are standard, {\em local} transformations
of the wave function, implied by the necessary rearrangements of the
phase-space coordinates. In the present paper we are going to show that the
above construction does not depend upon specific choice of coordinates used in
the above example but has a deep geometric meaning. This way not only harmonic
oscillator, but any linear quantum system evolves according to a similar law.
In fact, formula \eqref{eq:FrFT} is a special case of kernels which arise in a
natural way in geometric quantization \cite{Gaw, Kij3}, whenever we want to
describe transformation between two representations of a quantum state. All
these kernels may be defined in a geometric, coordinate-independent way. They
arise as superpositions of two standard building blocks: 1) the (appropriately
geometrized) Fourier transformation and 2) the generalized Galilei
transformation. To prove this fact, we analyze in the next Section the
geometric structure of a~quantum-mechanical wave function in terms of the
phase-space geometry.

\section{Geometric quantization}
\label{sec:GQ}

Consider the classical phase space $(\mathcal{P},\Omega)$ of a
system with $n$-degrees of freedom. This means that $\dim
\mathcal{P} = 2n$. By $\Omega$ we denote the canonical symplectic
form. Locally, a coordinate system $(x^i,p_i)$, $i = 1, \dots ,
n$, may be found, such that $\Omega$ reduces to the following
expression:
\begin{equation}\label{canonical}
    \Omega=\d p_i\wedge\d x^i \ ,
\end{equation}
where the summation convention is always used. Such coordinates
are called ``canonical coordinates''. In this paper we consider
the simplest, topologically trivial case $\mathcal{P} \simeq
\mathbb{R}^{2n}$, when canonical coordinates exist globally.

As a consequence of the Heisenberg uncertainty relation,
quantum-me\-cha\-nical wave function can not depend upon all these
phase-space coordinates but only upon a half of them. Physically,
this means that a representation of quantum states in terms of
wave functions is possible only {\em with respect} to a ``complete
system of commuting observables''. Examples, such as the
``position representation'' (wave functions depend upon position
variables $(x^i)$) or the ``momentum representation'' (wave
functions depend upon momenta $(p_i)$) are well known.
Geometrically, a ``system of commuting observables'' may be
considered as a foliation $\Lambda$ of $\mathcal{P}$ by the
congruence of all $n$-dimensional surfaces $\{(x^i,p_i):
x^i=\mbox{\rm const.}\}$ for the position representation and
surfaces $\{(x^i,p_i): p_i=\mbox{\rm const.}\}$ for the momentum
representation, respectively. The leaves of the above foliations
are {\em Lagrangian} submanifolds of $\mathcal{P}$. This means
that: 1) they are isotropic with respect to the canonical 2-form
\eqref{canonical} and 2) they have maximal dimension which is
possible for isotropic surfaces, namely a half of the dimension of
$\mathcal{P}$. Geometric quantization is a~theory which describes
the intrinsic properties of the quantum state in a~geometric,
coordinate-independent way.

\subsection{Quantum states and generalized Galilei transformation}
\label{sec:QS}

To give geometric definition of a quantum state, the following
three observations have to be taken into account:

\begin{figure}[htbp]
    \centering
    \psfrag{L}{$\Lambda$}
    \psfrag{P/L}{$Q_\Lambda=\mathcal{P}/\Lambda$}
    \includegraphics[width=0.35\textwidth]{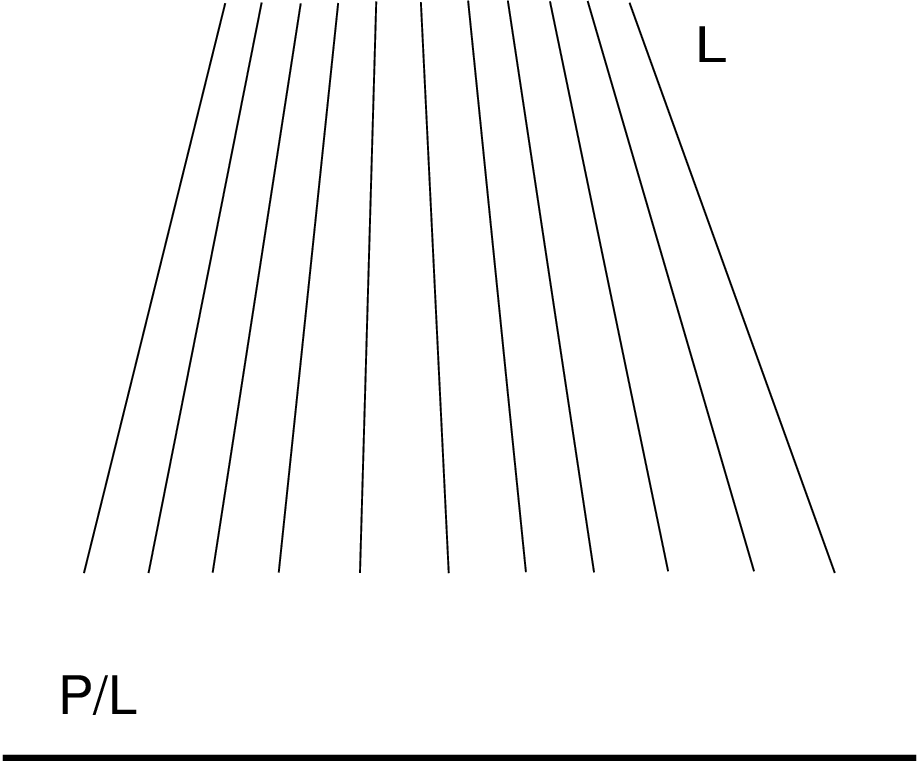}
    \label{fig:Q}
\end{figure}
1) Whenever a global, Lagrangian foliation $\Lambda$ of the phase
space has been chosen, the space of fibers
\begin{equation}\label{Qu}
    Q_\Lambda=\mathcal{P}/\Lambda
\end{equation}
plays the role of a generalized configuration space. Physically, it
describes independent variables (control parameters) of the
system. The wave function is an object living on $Q_\Lambda$.
Description of a quantum state {\em via} such a wave function will
be called the $Q_\Lambda$-representation. As an example we can
take the position or the momentum representation at different
instants $t$ of time.

2) To be able to calculate probabilities or transition amplitudes,
we have to integrate over the configuration space $Q_\Lambda$. For
this purpose people usually assume that a measure $\rho$ on
$Q_\Lambda$ has been chosen, such that the transition probability
between $\psi_1$ and $\psi_2$ is given by their scalar product in
$L^2(Q_\Lambda, \rho)$, namely:
\begin{equation}\label{product}
    (\psi_1 | \psi_2):= \int_{Q_\Lambda} \psi_1^* \psi_2\,
    \mbox{\rm d} \rho \ .
\end{equation}
Usually one chooses the Lebesgue measure carried by any system of
coordinates on the configuration space. Unfortunately, such a
description depends upon an arbitrary choice of coordinates. Even a change of
units (i.e.~centimeters {\em versus} inches) must be compensated by an
appropriate re-scaling of the wave function, cf.~formula \eqref{compensate}.
But only the quantity $\psi_1^* \psi_2\,  \mbox{\rm d} \rho $ has a physical
meaning. This quantity is independent upon all these (arbitrary) re-scalings.
An obvious simplification of the formalism consists in ``incorporating'' the
``square root of the measure'' into the wave
function. Namely, we consider the intrinsic half-density
\begin{equation}\label{half}
    \Psi := \psi \cdot \sqrt{{\rm d}\rho}
\end{equation}
(cf.~\cite{Abraham-Marsden}) instead of the scalar function $\psi$. This 
object does not depend upon any choice of coordinates nor the choice of any 
measure on $Q_\Lambda$. Assuming that the above
half-density is locally absolutely continuous with respect to the
square root of the Lebesgue measure carried by any system of
coordinates $(x^i)$ on $Q_\Lambda$, we may recover the
traditional, scalar wave function $\psi$ as the ratio between
$\Psi$ and the reference half-density $\sqrt{| \mbox{d} x^1 \wedge
\mbox{d} x^2 \wedge \cdots \wedge \mbox{d} x^n |}$. Complex
half-densities, square-integrable, absolutely continuous with
respect to Lebesgue, form a Hilbert space which will be called
$L^2(Q_\Lambda)$.

3) The above Hilbert space can not, however, be identified with
the physical space of states, because it does not reflect properly
the Galilei transformations of the wave function, due to the
change of the reference frame. Indeed, quantum representation of the momentum
$p_i$ canonically conjugate to the position $x^i$ is given by formula
\eqref{rep-pi}. We have seen in the previous Section that a generalized
Galilei transformation: $\tilde{p}_i = p_i +  \frac{\partial}{\partial x^i}
S(x)$, must be implemented on the level of quantum mechanics by the
multiplication of the wave function by the phase factor, as in formula
\eqref{GAL-Quantum}.

Observe that the above Galilei transformation consists in shifting
the value of $p$ by a constant value $\frac{\partial}{\partial
x^i} S(x)$ in each fiber $q \in Q_\Lambda$ independently. In order
to choose a specific one among all the possible canonical momenta
$p_i$, we have to choose at each fiber $q \in Q_\Lambda$ the point
where this observable vanishes. The collection of all these points
forms a Lagrangian surface $\lambda := \{ p_i = 0 \} \subset {\cal
P}$ which is transversal with respect to the foliation $\Lambda$.
We conclude that a choice of such a surface corresponds to a
choice of a reference frame.

We are going to show in the next Section that each fiber $q \in Q_\Lambda$
carries a natural {\em affine} structure. Choosing a specific ``reference
point'' $\lambda \cap q \in q$ transforms it into a {\em vector} space.
Moreover, we prove that this vector space (i.e.~space tangent to the fiber
$q$) is canonically equivalent to the cotangent space $T_q^*Q_\Lambda$. Choice
of a reference frame implies, therefore, that the fibration $\Lambda$ acquires
the vector-bundle structure isomorphic to the cotangent bundle $T^*Q_\Lambda$.
\begin{figure}[htbp]
    \centering
    \psfrag{L}{$\Lambda$}
    \psfrag{Q}{$Q_\Lambda$}
    \psfrag{a}{$\tilde\lambda$}
    \psfrag{b}{$\lambda$}
    \psfrag{q}{$q$}
    \includegraphics[width=0.45\textwidth]{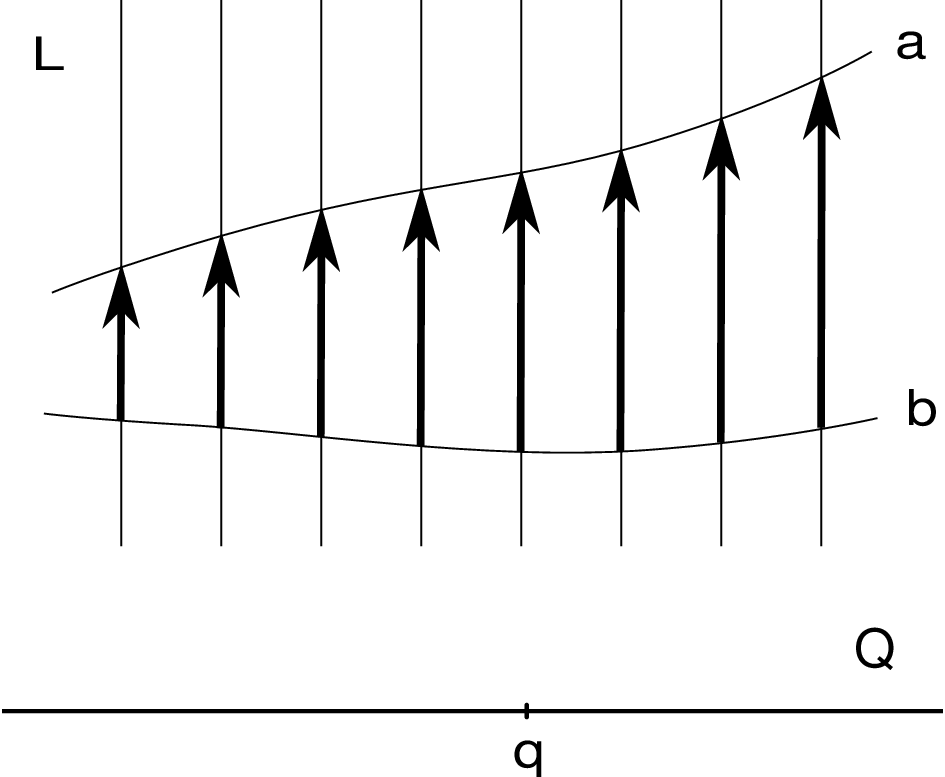}
    \label{fig:lambda}
\end{figure}

Suppose now that another Lagrangian surface $\tilde{\lambda} := \{ \tilde{p}_i
= 0 \} \subset {\cal P}$ (i.e.~another reference frame) has been chosen. The
difference between the two points: $\tilde{\lambda} \cap q$ and
$\lambda \cap q$,
defines in each affine space $q$ a tangent vector or, equivalently, a covector
on $Q_\Lambda$ attached at $q$. The collection of these covectors forms a
differential 1-form on the configuration space $Q_\Lambda$, which we denote by
$\tilde{\lambda} - \lambda$. As a consequence of the fact that both surfaces
$\tilde{\lambda}$ and $\lambda$ were Lagrangian we obtain an obvious

{\bf Corollary}: The form $\tilde{\lambda} - \lambda$ is closed.

Due to our topological assumption we have, therefore:
\begin{equation}
    \tilde{\lambda} - \lambda = \mbox{\rm d}
    S_{\tilde{\lambda}, \lambda } \ ,
\end{equation}
which, otherwise, would be true only locally. The function
$S_{\tilde{\lambda}, \lambda }$ is defined up to an additive constant.

As we have already discussed in the previous Sections, the elementary quantum
mechanics implies that the wave functions
describing the same quantum state with respect to different reference frames:
$\lambda := \{ p_i = 0 \}$ and $\tilde\lambda
:= \{ \tilde{p}_i = 0 \}$, differ by a phase factor, namely:
\begin{equation}\label{Gal-wf-2}
    \Psi_{\Lambda, \lambda}=\Psi_{\Lambda, \tilde{\lambda}} \cdot
      \e^{\frac {\i}{\hbar}  S_{\tilde{\lambda} , \lambda} } \ ,
\end{equation}
where the function $S_{\tilde{\lambda}, \lambda }$ is uniquely (up
to an additive constant) defined by the two submanifolds:
$\lambda$ and $\tilde{\lambda}$.

We see that to assign a wave function to a quantum (pure) state, it is not
sufficient to fix a ``complete system of commuting
observables'' (i.e. a~foliation $\Lambda$) but it is also necessary to choose
a reference frame (i.e. a~Lagrangian surface transversal to $\Lambda$). The
same quantum state, within the same representation $\Lambda$ (i.e.~position or
momentum representation) is represented by different wave functions with
respect to different reference frames. This suggests the following

{\bf Definition}: Quantum state in a representation $\Lambda$ is a class of
equivalent wave functions:
\[
  \mathfrak{Q}_{\Lambda} := [\Psi_{\Lambda,\lambda} ]
\]
where the equivalence relation is given by the generalized Galilei
transformation:
\begin{equation}\label{eq:GGT}
  \Psi_{\Lambda,\lambda} \sim
  \Psi_{\Lambda,\tilde{\lambda}} \Longleftrightarrow
  \left\{
  \Psi_{\Lambda,\lambda}=\Psi_{\Lambda,\tilde{\lambda}} \cdot
  \e^{\frac{\i}{\hbar} S_{\tilde{\lambda},\lambda}}\ \ ;\quad
  \d S_{\tilde{\lambda},\lambda}=\tilde{\lambda}-\lambda \right\} \ .
\end{equation}
Observe that the space $\mathcal{H}_{\Lambda}$ composed of all
quantum states is a projective Hilbert space because a constant
phase factor $\e^{\i c}$, $c\in\mathbb{R}$, of the wave function
(i.~e.~an additive constant of $S_{\tilde{\lambda},\lambda}$) is always
out of control.

\subsection{Proof of the affine-bundle structure of a Lagrangian foliation }
\label{sec:VBStr}

The coordinate-free construction of the affine-bundle structure of $\Lambda$
and of the phase function $S_{\tilde{\lambda}, \lambda
}$ in terms of the phase-space geometry was given in \cite{Kij3}.
It may be briefly sketched as follows:

\begin{figure}[htbp]
    \centering
    \psfrag{L}{$\Lambda$}
    \psfrag{Q}{$Q_\Lambda$}
    \psfrag{b}{$\kappa$}
    \psfrag{a}{$\kappa'$}
    \psfrag{q}{$q$}
    \psfrag{pp}{$\vect{p}'$}
    \psfrag{p1}{$\vect{p}_1$}
    \psfrag{p2}{$\vect{p}_2$}
    \psfrag{P}{$\vect{P}$}
    \includegraphics[width=0.45\textwidth]{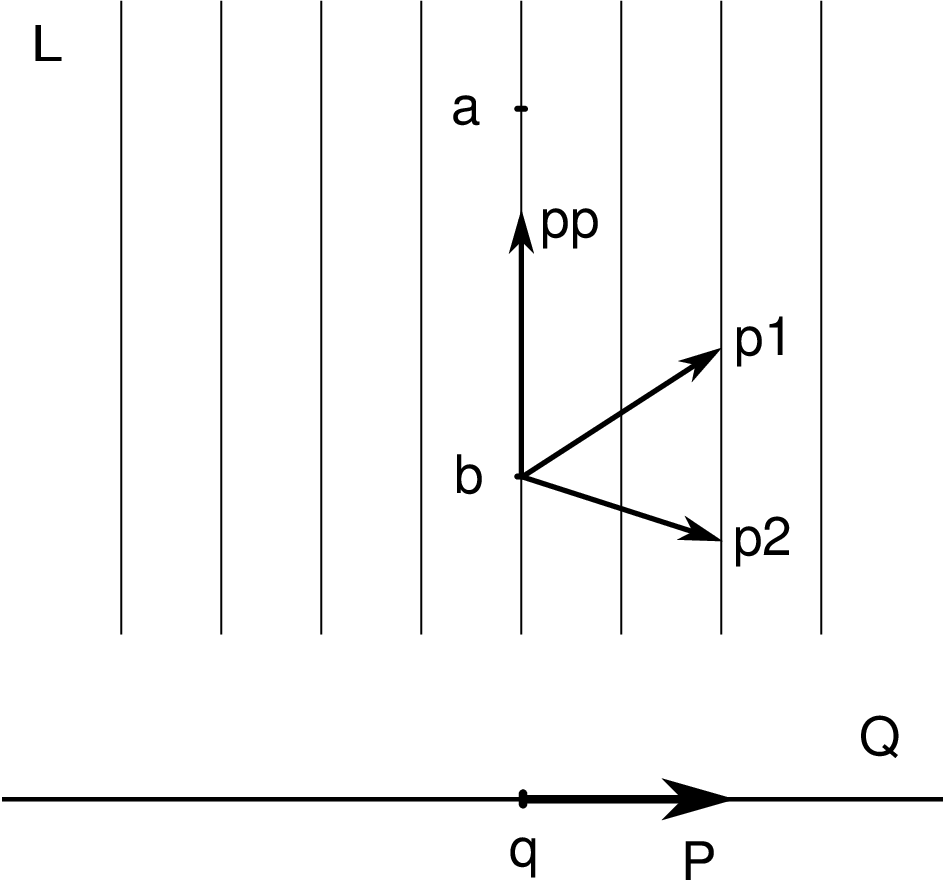}
    \label{fig:F_kappa}
\end{figure}
Given a fiber $q \in \Lambda$ and a point on it, $\kappa \in q$,
vectors tangent to $q$ at $\kappa$ can be canonically identified
with covectors on $Q_\Lambda$, attached at~$q$. The identification
is given by the formula:
\begin{equation}\label{isomorphism}
    \langle  \vect{P} | \vect{p}^\prime \rangle :=
   \Omega (\vect{p} , \vect{p}^\prime )
\end{equation}
Here, $\vect{P} \in T_q Q_\Lambda$ is a vector tangent to
$Q_\Lambda$ at $q$ and $\vect{p}^\prime \in T_{\kappa} q \subset
T_\kappa \mathcal{P}$ is a vector tangent to the fiber $q$.
By $\vect{p} \in T_\kappa \mathcal{P}$ we denote any vector which
projects onto $\vect{P}$ with respect to the canonical projection
in the fiber bundle $\pi: \mathcal{P}\to Q_\Lambda$. The value of
$\vect{p}^\prime$ on $\vect{P}$ is, therefore, equal to its
``symplectic scalar product'' with any representative $\vect{p}$
of $\vect{P}$, i.e.~with a vector $\vect{p}$ fulfilling: $\pi_*
\vect{p} = \vect{P}$. Of course, such a vector is not unique.
But for any pair of such vectors, say
$\vect{p}_1$ and $\vect{p}_2$, their difference projects on zero,
i.~e.~must be tangent to the fiber $q$.
Consequently, we have:
\[
   \Omega (\vect{p}_1 - \vect{p}_2 , \vect{p}^\prime ) = 0
\]
because both $\vect{p}_1 - \vect{p}_2$ and $\vect{p}^\prime$ are
tangent to $q$ which is Lagrangian. This proves that the left hand
side of \eqref{isomorphism} is defined uniquely. This way we have
constructed a mapping
\[
   F_\kappa : T_{\kappa} q \to  T^*_q Q_\Lambda \ .
\]
The non-degeneracy of $\Omega$ implies that $F_\kappa$ is an
isomorphism.

The above construction defines an auto-parallelism (a flat
connection) on every fiber $q \in \Lambda$. Indeed, given two
points $\kappa , \kappa^\prime \in q$, their tangent spaces
$T_\kappa q$ and $T_{\kappa^\prime} q$ are canonically isomorphic
to the same cotangent space $T^*_q Q_\Lambda$ and, therefore, may
be canonically identified. Moreover, it is easy to check that vector fields
which are constant along fibers, are Hamiltonian vector fields generated by
functions on $Q_\Lambda$ (i.e.~functions constant on fibers of $\Lambda$). But
the Poisson bracket of two such functions vanishes identically. This proves
that constant vector fields do commute, i.e.~the connection is flat and 
torsion-free. We conclude that every fiber $q \in \Lambda$ may be treated as a 
subset of an affine space. For pedagogical reasons we assume in the sequel 
that the topology of the fibration is trivial, i.e.~the fiber covers the 
entire affine space.

Now, we are going to assign to every pair $( \tilde{\lambda}
,\lambda )$ of sections of the bundle $\mathcal{P} \to Q_\Lambda$ a
covector field on the configuration space $Q_\Lambda$. We denote
it by $\tilde{\lambda} - \lambda$. It is defined by the formula:
\begin{equation}\label{section}
    ( \tilde{\lambda} - \lambda )(q) :=
    \tilde{\lambda}\cap q - \lambda \cap q \ ,
\end{equation}
where the right hand side is a vector tangent to the fiber $q$,
connecting the two points, i.e.~a covector on $Q_\Lambda$.
Because both sections are Lagrangian submanifolds, the resulting
form is closed:
\[
   \mbox{\rm d} (\tilde{\lambda} - \lambda) = 0 \ .
\]
Hence, locally, it satisfies:
\[
    \tilde{\lambda} - \lambda = \mbox{\rm d}
    S_{\tilde{\lambda}, \lambda } \ .
\]
Due to the trivial topology of $Q_\Lambda$, the potential
$S_{\tilde{\lambda}, \lambda }$ exists globally and is defined
uniquely up to an additive constant.

\subsection{Generalized Fourier transformation}
\label{sec:GFT}

The only arbitrary element which remains in the description of a
quantum state is the Lagrangian foliation $\Lambda$, representing
a complete set of commuting observables. Now, we are going to
describe the transformation which undergoes the wave function of a
given quantum state when we pass from one foliation to the other.
This will cover i.~g.~transformation from the position to the
momentum representation. But, we may also consider two foliations
corresponding to the position representation $\Lambda(t) := \{
x(t) =\mbox{\rm const.} \}$ at two different instants of time:
$t_1$ and $t_2$. The transformation between these two foliations
represents quantum dynamics.

Assume, therefore, that we have two different foliations
$\Lambda_1$ and $\Lambda_2$ of the symplectic space $\mathcal{P}$.
We are going to define the transformation from
$\mathcal{H}_{\Lambda_1}$ to $\mathcal{H}_{\Lambda_2}$
\begin{equation}\label{trans}
    \mathcal{F}_{\Lambda_2\Lambda_1} \colon \mathcal{H}_{\Lambda_1}
    \to \mathcal{H}_{\Lambda_2}.
\end{equation}
as an integral operator acting on corresponding wave functions
(cf. \cite{Gaw}, \cite{Kij3}). Here, we limit ourselves to a
simplified version, which works for {\em transversal} foliations. This 
assumption means that any fiber $\lambda_1 \in \Lambda_1$ has a unique 
intersection point with any fiber $\lambda_2 \in \Lambda_2$.
In this case a fiber $\lambda_1 \in \Lambda_1$ defines a
reference frame for the description of a quantum state with
respect to $\Lambda_2$ and {\em vice versa}. 
\begin{figure}[htbp]
    \centering
    \psfrag{L1}{$\Lambda_1$}
    \psfrag{L2}{$\Lambda_2$}
    \psfrag{l2}{$\lambda_2$}
    \psfrag{l1}{$\lambda_1$}
    \psfrag{Q1}{$Q_{\Lambda_1}$}
    \psfrag{Q2}{$Q_{\Lambda_2}$}
    \psfrag{q1}{$q_1$}
    \psfrag{q2}{$q_2$}
    \psfrag{A}{$A$}
    \psfrag{B}{$B$}
    \psfrag{C}{$C$}
    \psfrag{D}{$D$}
    \includegraphics[width=0.6\textwidth]{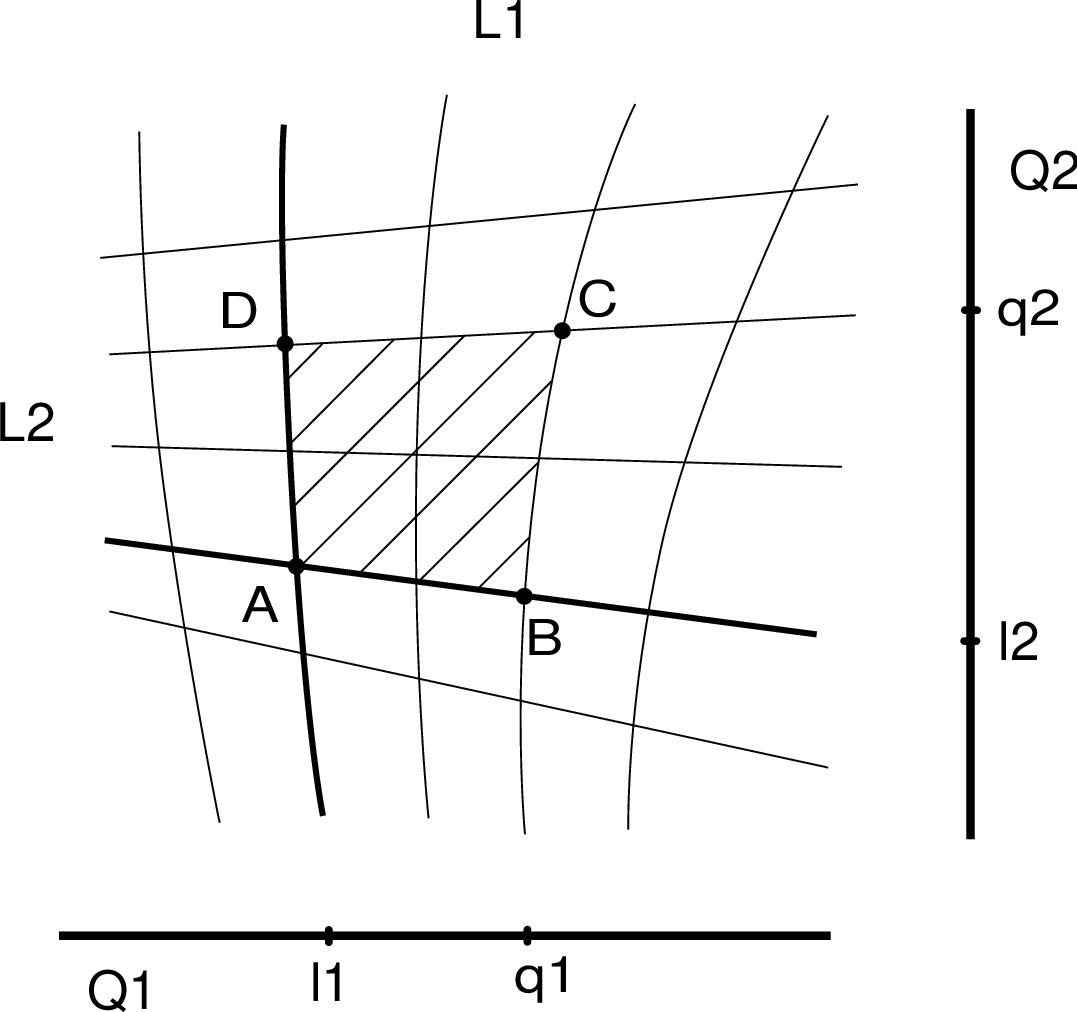}
    \label{fig:fourier}
\end{figure}
Choose, therefore, a pair $(\lambda_1 , \lambda_2)$, $\lambda_i \in 
\Lambda_i$; of such reference frames. Now, for any other pair $(q_1 , q_2)$, 
$q_i \in \Lambda_i$; consider the four intersection points: 1) $A=\lambda_1
\cap \lambda_2$, 2) $B= \lambda_2 \cap q_1$, 3) $C=q_1 \cap q_2$
and, finally, 4) $D= q_2 \cap \lambda_1$. Because every fiber
carries an affine structure, every pair of subsequent intersection
points defines uniquely an interval of a ``straight line''
connecting them (e.g.~we connect $A$ with $B$ along a straight
line in $\lambda_2$ and so on). This way we obtain uniquely an
oriented ``rectangle'' $ABCD$ which will be denoted $(\lambda_2,
q_1, q_2, \lambda_1)$, where the orientation is defined by the
sequence $(ABCD)$.
Define its ``symplectic surface''
$k(\lambda_2, q_1, q_2, \lambda_1)$ by:
\begin{equation}
  \label{surface}
    k(\lambda_2, q_1, q_2, \lambda_1):=\int_S \Omega
\end{equation}
where $S$ is any (oriented) 2-surface spanned by the rectangle,
i.e.~satisfying the condition: $\partial S =(\lambda_2, q_1, q_2,
\lambda_1)$. The definition does not depend upon a choice of such a
surface because the symplectic form $\Omega$ is closed. Indeed, if
$S_1$ and $S_2$ are two such surfaces, then there is a 3-volume
$V$ such that $\partial V = S_2 - S_1$ and, consequently, we have:
\[
    \int_{S_2} \Omega - \int_{S_1} \Omega = \int_{\partial V} \Omega
    = \int_V \mbox{d} \Omega =0 \ .
\]
The mapping \eqref{trans} is defined as the integral
transformation of the corresponding wave functions:
\begin{equation}\label{eq:GFT}
    \Psi_{\Lambda_2, \lambda_1}(q_2)
    = \int_{Q_{\Lambda_1}} \Psi_{\Lambda_1, \lambda_2}(q_1)
    K_{\lambda_1, \lambda_2}(q_1, q_2) \ ,
\end{equation}
where the kernel $K$ is defined as follows:
\begin{equation}\label{eq:K}
    K_{\lambda_1, \lambda_2}(q_1, q_2) =
    \sqrt{\big|\big(\tfrac{1}{\i\hbar}\Omega\big)^n\big|}
     \cdot \e^{-\frac{\i}{\hbar} k(\lambda_2, q_1, q_2, \lambda_1)}
      \ .
\end{equation}
Here, $2n=\dim \mathcal{P}$,
$\Omega^n=\Omega\wedge\Omega\wedge\dots\wedge\Omega$ is a $2n$-form
(scalar density) on $\mathcal{P}$, $\sqrt{| \Omega^n |}$ is the
corresponding half-density.

If $(x^i)$, $i=1, \dots , n$, are coordinates on $Q_{\Lambda_1}$
and $(y^i)$, $i=1, \dots , n$, are coordinates on $Q_{\Lambda_2}$,
then $(x^i , y^j)$ define a coordinate chart on the phase space
$\mathcal{P}$. Hence, the $2n$-form $\Omega^n $ is proportional to
$\mbox{d}x^1 \wedge \cdots \wedge \mbox{d}x^n \wedge \mbox{d}y^1
\wedge \cdots \wedge \mbox{d}y^n$. Consequently, we have:
\begin{equation}\label{Liu}
    \sqrt{| \Omega^n |} = f(x,y)
    \sqrt{| \mbox{d}x^1 \wedge \cdots \wedge \mbox{d}x^n |}
    \sqrt{| \mbox{d}y^1 \wedge \cdots \wedge \mbox{d}y^n |}
\end{equation}
Because wave function $\Psi_{\Lambda_1, \lambda_2}$ is a
half-density on $Q_{\Lambda_1}$, it contains already the factor
$\sqrt{| \mbox{d}x^1 \wedge \cdots \wedge \mbox{d}x^n |}$.
Together with the same factor from \eqref{Liu} it produces the
scalar density on $Q_{\Lambda_1}$ which we integrate according to
formula \eqref{eq:GFT}. The result of this integration contains
the remaining factor $\sqrt{| \mbox{d}y^1 \wedge \cdots \wedge
\mbox{d}y^n |}$ from \eqref{Liu}, i.e.~produces a half-density on
$Q_{\Lambda_2}$.

The operator $\mathcal{F}_{\Lambda_2\Lambda_1}$ is called the
generalized Fourier transformation. It is well defined for any
pair of transversal foliations. In the present paper we shall use
it thoroughly in a specific case, when the two foliations are {\em
compatible}. It turns out that this covers all the cases of {\em
linear dynamics} (e.g.~free motion, harmonic oscillator and a
constant electric or magnetic fields). As will be seen in the next
Section, the entire information about the quantum dynamics can be
obtained from its classical counterpart by means of the operator
$\mathcal{F}_{\Lambda_2\Lambda_1}$.

For the sake of completeness we shall now formulate the
compatibility condition, which implies specific properties of the
generalized Fourier kernel \eqref{eq:K}. For this purpose observe
that, given two transversal foliations, there is a unique and
natural way to transport vectors tangent to fibers of $\Lambda_1$
along the fibers of $\Lambda_2$. Indeed, given a fiber $\sigma \in
\Lambda_2$, two vectors $\vect{p}$ and $\vect{r}$, tangent to
$\lambda \in \Lambda_1$ and $q \in \Lambda_1$ at the points
$\lambda \cap \sigma$ and $q \cap \sigma$ respectively, may be
identified if they project onto the same vector tangent to
$Q_{\Lambda_2}$, i.e.~if $\pi_* \vect{p} = \pi_* \vect{r}$.
\begin{figure}[htbp]
    \centering
    \psfrag{L1}{$\Lambda_1$}
    \psfrag{L2}{$\Lambda_2$}
    \psfrag{Q1}{$Q_{\Lambda_1}$}
    \psfrag{Q2}{$Q_{\Lambda_2}$}
    \psfrag{q}{$q$}
    \psfrag{l}{$\lambda$}
    \psfrag{s}{$\sigma$}
    \psfrag{p}{$\vect{p}$}
    \psfrag{r}{$\vect{r}$}
    \psfrag{P}{$\pi_* \vect{p} = \pi_* \vect{r}$}
    \includegraphics[width=0.5\textwidth]{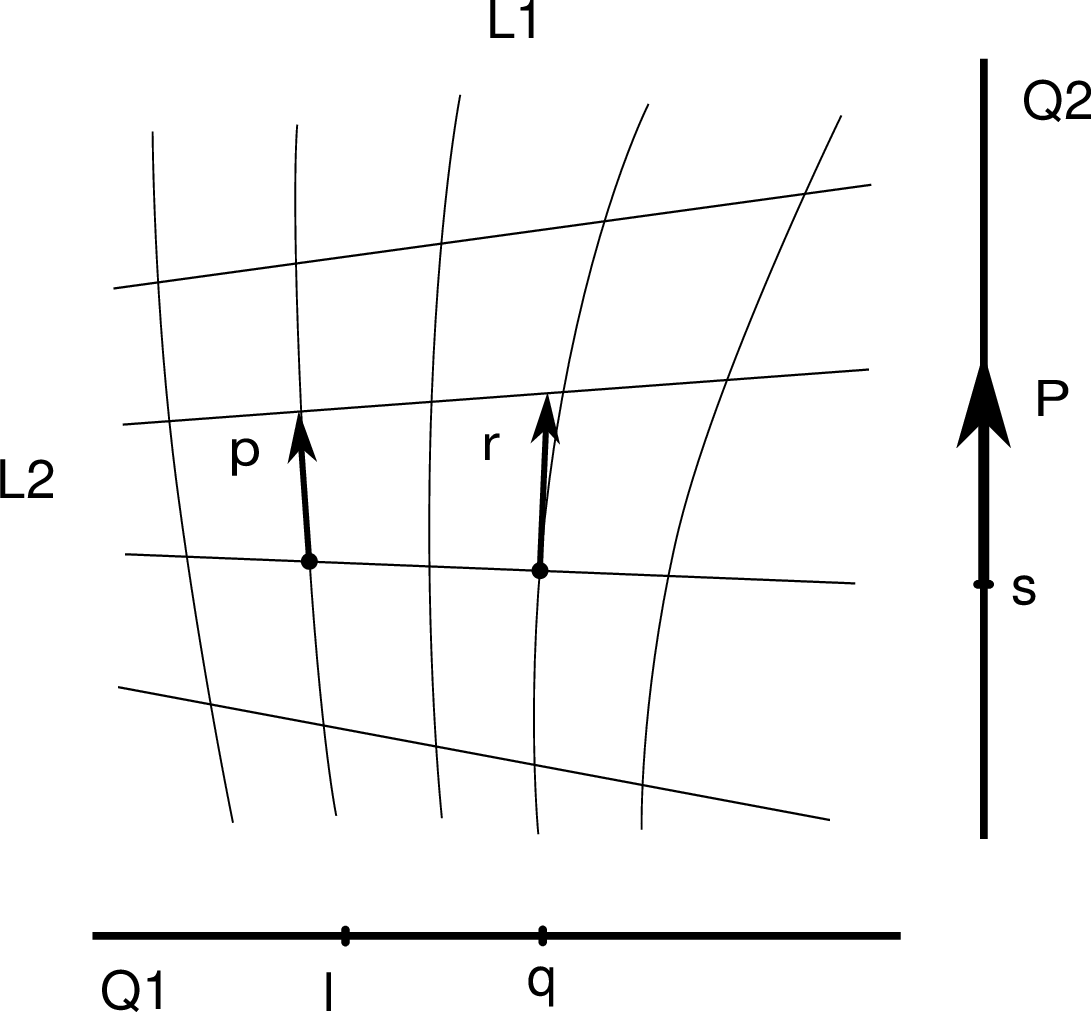}
    \label{fig:transport}
\end{figure}

On the other hand, the affine structure of the fibers allows us to
transport them parallelly along fibers of $\Lambda_1$.

{\bf Definition}: Two mutually transversal, Lagrangian foliations
$\Lambda_1$ and $\Lambda_2$ are called {\em compatible} if
parallel translations along $\Lambda_1$ commute with those along
$\Lambda_2$.

An obvious example of compatible foliations is given by the
position foliation $\{ x^i = \mbox{const.} \}$ and the momentum
foliation $\{ p_i = \mbox{const.} \}$, if $(x^i , p_i)$ are
canonical variables. In case of compatible foliations the operator
$\mathcal{F}_{\Lambda_2\Lambda_1}$ is unitary and fulfills the
chain rule: $\mathcal{F}_{\Lambda_3\Lambda_2}
\mathcal{F}_{\Lambda_2\Lambda_1}=\mathcal{F}_{\Lambda_3\Lambda_1}$.

The proof of this property may be sketched as follows. For {\em
compatible} foliations the function $f$ in formula \eqref{Liu}
factorizes and we have: $f(x,y) = h(x) \cdot k(y)$. On the other
hand, both $Q_{\Lambda_1}$ and $Q_{\Lambda_2}$ carry an affine
structure and the corresponding vector spaces are in canonical
duality. It is easy to see that the phase factor $k(\lambda_2,
q_1, q_2, \lambda_1)$ is given by the above duality form
\begin{equation}\label{duality}
    k(\lambda_2, q_1, q_2, \lambda_1) :=
    \big\langle (q_1 - \lambda_1)(\lambda_2) \big|
    (q_2 - \lambda_2)(\lambda_1) \big\rangle \ .
\end{equation}
Hence, the entire kernel \eqref{eq:K} factorizes and reduces to
the standard Fourier kernel written in linear coordinates
compatible with the affine structure carried by the two
foliations. This implies the group properties of the
transformation.

\section{Symmetries between linear quantum systems}
\label{sec:LD}

We stress that there was no linear structure of the configuration
or the phase spaces assumed {\em a priori}. The symplectic form
implies the affine structure of the fibers of the Lagrange'an
foliation $\Lambda$. However, if we take two {\em compatible}
foliations $\Lambda_1$ and $\Lambda_2$, then the entire phase
space $\cal P$ acquires an affine structure.

In this context the {\em linear} dynamics has to be understood as
a specific situation, for which the ``position-foliations''
$\Lambda_t :=\{ x(t) = \mbox{const.} \}$ remain mutually
compatible for different times $t_1$ and $t_2$. It is easy to
check that this happens if and only if there are canonical
variables in $\cal P$, such that the Hamiltonian is at most
quadratic.

In this section we analyze examples of linear dynamics in the
geometric quantization context. We prove that our generalized
Fourier transformation gives the correct quantum evolution. We
begin with the classical analysis which shows that the
configuration foliations $\Lambda_t :=\{ x(t) = \mbox{const.} \}$
are, in fact, the same for all possible cases of linear dynamics.
This implies that any solution of the Schr\"odinger equation with
at most quadratic potential (e.g.~harmonic oscillator, constant
electric or magnetic fields) is uniquely given by a corresponding
solution describing the free motion. We conclude that different
linear quantum systems are, essentially, all the same.

\subsection{Harmonic oscillator vs. free motion}
\label{sec:HarmonicOscilatorVsFreeMotion}

Consider classical dynamics of a free particle (for simplicity we
limit ourselves to 1 degree of freedom)\footnote{Hamiltonian
equations for free particle:
\begin{equation*}
    \dot{\vect{r}}(t) = \frac 1 m \vect{p}(t), \qquad \dot{\vect{p}}(t) = 0.
\end{equation*}
}:
\begin{equation}\label{eq:free_solve}
  \begin{split}
      x(t) &= x(0) + \frac{t}{m}p(0),\\
      p(t) &= p(0),
  \end{split}
\end{equation}
and of a harmonic oscillator\footnote{Hamiltonian equations for
harmonic oscillator:
\begin{equation*}
    \dot{\tilde{\vect{r}}}(\tau) = \frac 1 m \tilde{\vect{p}}(\tau),
    \qquad \dot{\tilde{\vect{p}}}(\tau) = -k \tilde{\vect{r}}(\tau).
\end{equation*}
}:
\begin{equation}\label{eq:osc_solve}
  \begin{split}
      \tilde{x}(\tau) &= \cos\omega\tau\, \tilde{x}(0)
      + \frac{1}{m\omega} \sin\omega\tau\, \tilde{p}(0),
      \\
      \tilde{p}(\tau) &= - m\omega  \sin\omega\tau\, \tilde{x}(0)
      + \cos\omega\tau\, \tilde{p}(0),
  \end{split}
\end{equation}
where $\omega=\sqrt{\frac{k}{m}}$. Suppose that positions and
momenta coincide at initial time $t=0$ (i.e.~we have:
$\{x(0)=\mbox{const.}\}=\{\tilde{x}(0)=\mbox{const.}\}$ and
$\{p(0)=\mbox{const.}\}=\{\tilde{p}(0)=\mbox{const.}\}$). Observe that for
\[
  t = \frac{1}{\omega} \tan\omega\tau,
\]
the configuration foliations of the two systems coincide.
\begin{figure}[htbp]
    \centering
    \psfrag{x}{$\{x(0)=\mbox{const.}\}=\{\tilde{x}(0)=\mbox{const.}\}$}
    \psfrag{p}{$\{p(0)=\mbox{const.}\}=\{\tilde{p}(0)=\mbox{const.}\}$}
    \psfrag{p0}{$p(0)=0$}
    \psfrag{pt0}{$\tilde{p}(\tau)=0$}
    \psfrag{xt}{$\{x(t)=\mbox{const.}\}=\{\tilde{x}(\tau)=\mbox{const.}\}$}
    \includegraphics[width=0.6\textwidth]{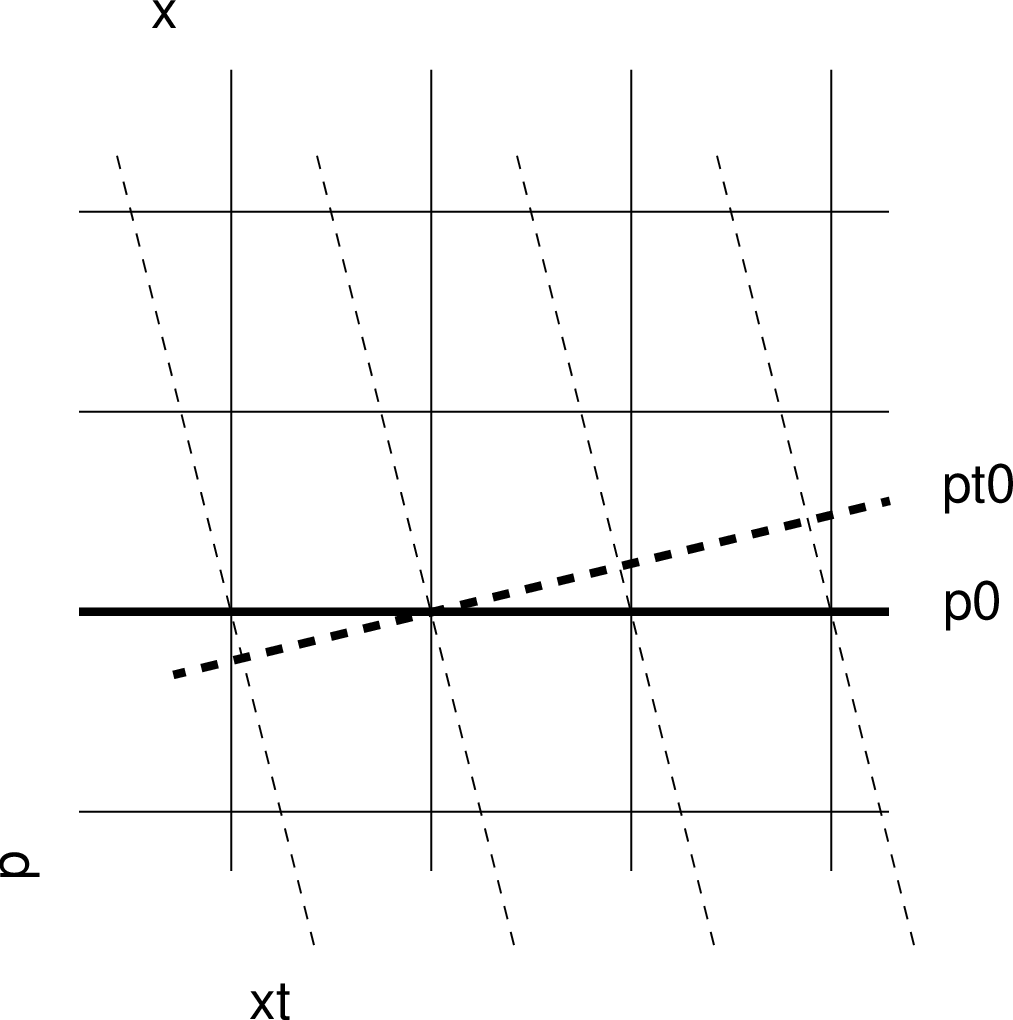}
    \label{fig:xp}
\end{figure}
Indeed, we have:
\[
  \sin \omega\tau = \frac{\omega t}{\sqrt{1+\omega^2 t^2}}, \qquad
  \cos \omega\tau = \frac{1}{\sqrt{1+\omega^2 t^2}},
\]
and, therefore, equations \eqref{eq:osc_solve} imply the following
relations:
\begin{align}
    \tilde{x}(\tau) &= \frac{1}{\sqrt{1+\omega^2 t^2}}\,x(t),\\
    \tilde{p}(\tau) &= - \frac{m\omega^2 t}{\sqrt{1+\omega^2 t^2}}\, x(t)
      + \sqrt{1+\omega^2 t^2}\, p(t). \label{odn}
\end{align}
This proves that foliations $\{ x(t) = \mbox{const.} \}$ and $\{
\tilde{x}(\tau) = \mbox{const.} \}$ do coincide. If, therefore,
$\phi (t,x)\sqrt{\d x}$ is the wave function of a free moving
particle, and $\psi({\tau, \tilde{x}})\sqrt{\d\tilde{x}}$ is a
wave function of a harmonic oscillator, both starting with the
same initial condition at $t=0$, than both wave functions must
coincide up to a Galilei transformation. Such a transformation is
necessary because, due to \eqref{odn}, the reference section
$\sigma :=\{ \tilde{p}(\tau) = 0 \}$ for the harmonic oscillator
corresponds to
\[
   p(t) = \frac{m\omega^2 t}{1+\omega^2 t^2}\, x(t) \ .
\]
But the wave function of the free motion describes the quantum
state with respect to the reference $\lambda:= \{ p(t) = 0 \}$. We
have, therefore:
\[
   \sigma - \lambda = \frac{m\omega^2  t}{1+\omega^2 t^2}\, x
    = \d \left( \frac 12\, \frac{m\omega^2  t}{1+\omega^2 t^2}\, x^2
    \right) \ .
\]
The quantity in brackets describes the phase of the Galilei
factor. We conclude that the following equality holds:
\begin{equation}
\label{eq:phi=psi}
    \phi(t,x)\sqrt{\d x} = (1 + \omega^2 t^2)^{-\frac 1 4}
    \psi\Big(\frac{1}{\omega}\arctan{\omega t},
    \frac{x}{\sqrt{1+\omega^2 t^2}}\Big)
    \,\e^{\frac{\i}{2\hbar}\, \frac{m\omega^2 t}{1+\omega^2 t^2}\, x^2}
    \sqrt{\d x},
\end{equation}
because $\sqrt{\d\tilde{x}}=(1 + \omega^2 t^2)^{-\frac 1
4}\sqrt{\d x}$. It is easy to check that the function $\phi$
satisfies  the free Schr\"odinger equation if and only if $\psi$
satisfies the Schr\"odinger equation for the harmonic oscillator.

The above {\em local} transformation between solutions of the free
Schr\"odinger equation and the harmonic oscillator was first found
by U.~Niederer (see \cite{Nied1, Nied2}) and then reinterpreted by
A.~O.~Barut in terms of the conformal group representation (see
\cite{Barut}). In paper \cite{KijJak} the same symmetries were
derived as the only local symmetries in the multisymplectic
formulation of the Schr\"odinger equation.
The formula \eqref{eq:phi=psi} is also known as the \textit{lens
transform} in the field of nonlinear Schr\"odinger equations (see
\cite{Rybin, Carl1}).

\subsection{Motion in constant electric field vs. free motion}
\label{sec:ParticleInConstantElectricFieldVsFreeMotion}

A similar relation between free motion and the motion of a charged
particle in a constant electric (or gravitational) field\footnote{
Hamiltonian equations for a charged particle moving in a constant
electric field $\vect{E}$:
\begin{equation*}
    \dot{\tilde{\vect{r}}}(\tau) = \frac 1 m \tilde{\vect{p}}(\tau),
    \qquad \dot{\tilde{\vect{p}}}(\tau) = e\vect{E}.
\end{equation*}
} can be proved (again, for the maximal simplicity we limit
ourselves to the 1D case):
\begin{equation}\label{eq:el_solve}
  \begin{split}
    \tilde{x}(\tau) &= \tilde{x}(0) + \frac{\tau}{m} \tilde{p}(0)
    + \frac{eE\tau^2}{2m}\ ,\\
    \tilde{p}(\tau) &= \tilde{p}(0) + eE\tau \ .
  \end{split}
\end{equation}
We see that the position foliations $\{x(t)=\mbox{const.}\}$ and $\{
\tilde{x}(\tau)=\mbox{const.} \}$ coincide for $t = \tau $. Indeed,
comparing \eqref{eq:free_solve} with \eqref{eq:el_solve} we obtain
the following relations
\begin{align}
    \tilde{x}(\tau) &= x(t) + \frac{eEt^2}{2m},\nonumber \\
    \tilde{p}(\tau) &= p(t) + eEt. \label{ppp}
\end{align}
The last equation implies that the Galilei transformation from the
reference surface $\lambda:=\{ {p}(t)=0 \}$ to the new reference
surface $\sigma:= \{ \tilde{p}(\tau)=0 \}$ consists in a simple
shift by the constant (in variable $x$) value ``$eEt$''.
Consequently, the corresponding phase is linear in $x$ and equals
$eEtx$.

Let $\phi (t,x)\sqrt{\d x}$ be a wave function of a free particle,
and $\psi({\tau, \tilde{x}})\sqrt{\d\tilde{x}}$ -- a~wave function
of a charged particle, both starting from the same initial value
at $t=0$. Then, geometric quantization implies the following
identity:
\begin{equation}\label{eq:phi=psi_ele}
    \phi(t,x)\sqrt{\d x} = \psi\!\Big(t,x + \frac{eEt^2}{2m}\Big)
    \,\e^{-\frac{\i}{\hbar}eEtx}\sqrt{\d x},
\end{equation}
because $\sqrt{\d\tilde{x}}=\sqrt{\d x}$.

It is easy to check that the wave function $\phi$ satisfies the
free Schr\"odinger equation if and only if $\psi$ satisfies the
following Schr\"odinger equation:
\begin{align}
  \label{eq:Sch_const_E}
    \i\hbar \partial_{\tau} \psi(\tau,\tilde{x})
    = -\frac{\hbar^2}{2m} \partial^{2}_{\tilde{x}} \psi(\tau,\tilde{x})
    +\left( -eE\tilde{x} + \frac{e^2 E^2 \tau^2}{m} \right)
    \psi(\tau,\tilde{x})\;.
\end{align}
It describes the motion of our charged particle in the linear
potential:
\begin{equation}\label{linear-pot}
      U(\tilde{x}):= -eE\tilde{x} + \frac{e^2 E^2 \tau^2}{m} \ ,
\end{equation}
i.e. in the constant electric field. The non-vanishing additive
constant $\frac{e^2 E^2 \tau^2}{m}$ may be eliminated by the gauge
transformation
\begin{align}
  \label{eq:gauge}
    U \to U' = U + \partial_\tau \chi\;,
\end{align}
whereas the wave function transforms as follows:
\begin{align}
  \label{eq:wave_gauge}
    \psi'= \psi \e^{-\frac{\i}{\hbar} \chi} \ .
\end{align}
In our case we have
\begin{align}
    \chi = -\frac{e^2 E^2 \tau^3}{3m}\ .
\end{align}
We conclude that in the following equality:
\begin{equation}
  \phi(t,x)\sqrt{\d x} = \psi'\Big(t,x + \frac{eEt^2}{2m}\Big)
  \,\e^{-\frac{\i}{\hbar}eEtx}
  \e^{-\frac \i \hbar \frac{e^2 E^2 t^3}{3m}}\sqrt{\d x}
  \label{eq:free=const_E}
\end{equation}
the wave function $\phi$ satisfies the free Schr\"odinger equation
if and only if $\psi'$~satisfies the Schr\"odinger equation with
the standard potential
\[
  U'(\tilde{x}):= -eE\tilde{x}  \ .
\]
We stress, however, that these manipulations have no physical
meaning: both the additive constant in the potential and the
constant phase in the wave function have no physical
interpretation and cannot be controlled within the framework we
have defined in the present paper.

The transformation \eqref{eq:free=const_E} is also known in the literature
as the Avron-Herbst formula (see \cite{Carl2}).

\subsection{Motion in constant magnetic field vs. free motion}
\label{sec:ParticleInConstantMagneticFieldVsFreeMotion}

Finally, we are going to show a relation between solutions of the
Shr\"odinger equation for a free particle and for a charged
particle moving in a constant magnetic field.\footnote{In the case
of a constant magnetic field $\vect{B}$ we can choose the vector
potential in the following form: $\vect{A}=\frac 1 2
\vect{B}\times\tilde{\vect{r}}$. Then, the Hamiltonian $H =
\frac{(\tilde{\vect{p}}-e\vect{A})^2}{2m}$ implies the following
equations
\begin{equation*}
    \dot{\tilde{\vect{r}}}(\tau) = \frac{e}{2m}\tilde{\vect{r}}(\tau)
    \times \vect{B} + \frac 1 m \tilde{\vect{p}}(\tau)\;,
    \qquad
    \dot{\tilde{\vect{p}}}(\tau) = -\frac{e^2}{8m}
    \nabla (\vect{B}\times\tilde{\vect{r}}(\tau))^2 +
    \frac{e}{2m}\tilde{\vect{p}}(\tau)\times\vect{B}\;.
\end{equation*}
} Putting $\vect{B}=(0,0,B)$, we obtain the non-trivial motion in
the $(x,y)$ plane:
\begin{align}
  \label{eq:mag_solve_xy}
  \begin{split}
      \tilde{x}(\tau) &= \frac 1 2 (\cos \omega \tau + 1)\,\tilde{x}(0)
          + \frac 1 2 \sin \omega \tau\,\tilde{y}(0)\\
         &+ \frac{1}{\omega m}\sin \omega \tau\, \tilde{p}_{x}(0)
          - \frac{1}{\omega m}(\cos \omega \tau - 1)\,\tilde{p}_y (0), \\
      \tilde{y}(\tau) &= -\frac 1 2 \sin \omega \tau\, \tilde{x}(0)
          + \frac 1 2 (\cos \omega \tau + 1)\,\tilde{y}(0) \\
         &+ \frac{1}{\omega m}(\cos \omega \tau - 1)\,\tilde{p}_x (0)
          + \frac{1}{\omega m}\sin \omega \tau\, \tilde{p}_y (0),
  \end{split}
\end{align}
\begin{align}
  \label{eq:mag_solve_pxpy}
  \begin{split}
      \tilde{p}_x (\tau) &= -\frac 1 4 \omega m\sin\omega \tau\, \tilde{x}(0)
             + \frac 1 4 \omega m(\cos\omega \tau - 1)\,\tilde{y}(0) \\
         &+ \frac 1 2 (\cos \omega \tau + 1)\,\tilde{p}_x (0)
          + \frac{1}{2}\sin \omega \tau\, \tilde{p}_y (0),\\
      \tilde{p}_y (\tau) &= -\frac 1 4 \omega m(\cos \omega \tau
      - 1)\,\tilde{x}(0)
               - \frac 1 4 \omega m \sin \omega \tau\,\tilde{y}(0)\\
         &- \frac{1}{2}\sin \omega \tau\, \tilde{p}_{x}(0)
          + \frac{1}{2}(\cos \omega \tau + 1)\,\tilde{p}_y (0),
  \end{split}
\end{align}
where $\omega = \frac{eB}{m}$. Suppose, as usual, that the
appropriate foliations for the free motion and for the motion in
magnetic field overlap at the beginning:
$\big\{\big(x(0),y(0)\big)=\mbox{const.}\big\} =\big\{\big(\tilde{x}(0),
\tilde{y}(0)\big)=\mbox{const.}\big\}$ and
$\big\{\big(p_x(0),p_y(0)\big)=\mbox{const.}\big\}
=\big\{\big(\tilde{p}_x(0), \tilde{p}_y(0)\big)=\mbox{const.}\big\}$.
Then, the configuration foliations after time $t$ also overlap for
\[
  t = \frac{2\sin \omega\tau}{\omega(\cos \omega\tau + 1)}\;,
\]
or, equivalently,
\[
  \sin \omega\tau = \frac{4\omega t}{4+\omega^2 t^2}\ ,
  \qquad
  \cos \omega\tau = \frac{4-\omega^2 t^2}{4+\omega^2 t^2}\ .
\]
We obtain, therefore, relations:
\begin{align*}
    \tilde{x}(\tau) &= \frac{4}{4+\omega^2 t^2}\, x(t)
      + \frac{2\omega t}{4+\omega^2 t^2}\, y(t)\;,\\
    \tilde{y}(\tau) &= - \frac{2\omega t}{4+\omega^2 t^2}\, x(t)
      + \frac{4}{4+\omega^2 t^2}\, y(t)\;,
\end{align*}
\begin{align*}
    \tilde{p}_x(\tau) &= - \frac{m\omega^2 t}{2(4+\omega^2 t^2)}\, x(t)
    - \frac{m \omega^3 t^2}{2(4+\omega^2 t^2)}\, y(t) + p_x(t)
    + \frac 1 2 \omega t\, p_y(t)\;,\\
  \tilde{p}_y(\tau) &=  \frac{m\omega^3 t^2}{2(4+\omega^2 t^2)}\, x(t)
    - \frac{m \omega^2 t}{2(4+\omega^2 t^2)}\, y(t)
    - \frac 1 2 \omega t\, p_x(t) + p_y(t)\;.
\end{align*}

Suppose now that $\phi (t,x,y)\sqrt{\d x \d y}$ describes the free
particle evolution, and $\psi(\tau,
\tilde{x},\tilde{y})\sqrt{\d\tilde{x}\d\tilde{y}}$ the evolution
of a charged particle in the constant magnetic field. If both
functions satisfy the same initial conditions at $t = 0 = \tau$,
then we have the identity:
\begin{multline}\label{eq:phi=psi_mag}
    \phi(t,x,y)\sqrt{\d x \d y} = \sqrt{\frac{4}{4+\omega^2 t^2}} \\ \times
    \psi \Big(\frac{1}{\omega}\arcsin \frac{4\omega t}{4+\omega^2 t^2},
      \frac{4}{4+\omega^2 t^2}\, x
      + \frac{2\omega t}{4+\omega^2 t^2}\, y,
      - \frac{2\omega t}{4+\omega^2 t^2}\, x
      + \frac{4}{4+\omega^2 t^2}\, y \Big) \\ \times
      \e^{\frac{\i}{2\hbar}
      \frac{m\omega^2 t}{4 + \omega^2 t^2} (x^2+y^2)}\sqrt{\d x \d y}\;.
\end{multline}
The density factor comes from the identity:
$\sqrt{\d\tilde{x}\d\tilde{y}} =\sqrt{\frac{4}{4+\omega^2
t^2}}\sqrt{\d x \d y}$. The phase factor comes from the
generalized Galilei transformation \eqref{eq:GGT}, because
``zero'' surface $\{ {\vect{p}}(t)=0 \}$ has to be replaced by the
``new zero'': $\{ \tilde{\vect{p}}(\tau)=0 \}$. It is easy to
check that $\phi$ fulfills the free Schr\"odinger equation if and
only if $\psi$ fulfills the Schr\"odinger equation for a charged
particle moving in a constant magnetic field.

\section{Reproducing kernels for linear dynamics}
\label{sec:LD-solution}

Using our techniques we were able to show that any linear
evolution is isomorphic to the free evolution. For this purpose we
were not obliged to solve the Schr\"odinger equations. In the
present section we prove that also the solution of the initial
value problem can be easily obtained in terms of the generalized
Fourier and the Galilei transformations.

\subsection{Initial value problem for the free particle}
\label{sec:InitialProblemForFreeParticle}

Consider an initial quantum state at time $t=0$, which is
described by the wave function $\Psi_{\Lambda,\lambda}(x)$, where
$\Lambda$ is the corresponding configuration foliation, i.e. the
collection of all the fibers $\{(x,p): x=\mbox{const.}\}$, and
$\lambda=\{p=0\}$ is the ``zero'' (reference) surface. After the
lapse of the $t$, the classical evolution of the system leads to
the new variables $(x',p'):= (x(t), p(t))$. We are going to prove
that the corresponding quantum evolution leads {\em exactly} to
the wave function $\Psi_{\Lambda',\lambda'}(x')$, where $\Lambda'$
is the corresponding configuration foliation, i.e. the collection
of the fibers $\{(x',p'): x'=\mbox{const.}\}$, whereas
$\lambda'=\{p'=0\}$.

Indeed, the (purely ``static'') recalculation of the same quantum
state from the old representation $\Psi_{\Lambda,\lambda}(x)$ to
the new representation $\Psi_{\Lambda',\lambda'}(x')$ can be
performed in three steps:

\begin{enumerate}
    \item The generalized Galilei transformation between the reference
     $\lambda=\{p=0\}$ to the new reference $\mu'=\{x'=0\}$.
    Due to \eqref{eq:free_solve} we have:
		\[
		  (\lambda - \mu')(x)= \frac{m}{t}x \d x = \d\Big(\frac{m}{2t}x^2\Big),
		\]
		which determines (up to an additive constant) the phase function
		$S_{\mu',\lambda}=\frac{m}{2t}x^2$ . Hence, according to
		\eqref{eq:GGT}, we have:
		\begin{equation}\label{eq:mu'-lambda}
		  \Psi_{\Lambda,\mu'}(x)=\Psi_{\Lambda,\lambda}(x)\;
		  \e^{\frac{\i}{\hbar}\frac{m}{2t}x^2}.
		\end{equation}

    \item In  the second step, we perform the generalized
		Fourier transformation \eqref{eq:GFT}. Starting from the wave
		function $\Psi_{\Lambda\mu'}$ we obtain $\Psi_{\Lambda',\mu}$,
		where $\mu=\{x=0\}$:
		\begin{align}\label{eq:GFT_free}
		  \Psi_{\Lambda',\mu}(x') &= \int \Psi_{\Lambda,\mu'}(x)\;
		  \e^{-\frac{\i}{\hbar}\frac{m}{t}x'x}
		    \sqrt{\frac{m}{\i\hbar t}}\,\sqrt{\d x}\sqrt{\d x'},
		\end{align}
		because $\frac{1}{\i\hbar}\Omega=\frac{1}{\i\hbar} \d p\land \d
		x=\frac{m}{\i\hbar t}\d x'\land \d x$.

    \item  Finally, in order to calculate the wave function
    $\Psi_{\Lambda',\lambda'}$, we have to apply again
    the Galilei transformation
    from the reference $\mu=\{x=0\}$ to the new reference
    $\lambda'=\{p'=0\}$. Due to \eqref{eq:free_solve} we have
		\[
		  (\mu - \lambda')(x')= \frac{m}{t}x'\d x' = \d\Big(\frac{m}{2t}x'^2\Big).
		\]
		This determines the generating function $S_{\lambda',\mu}=
		\frac{m}{2t}x'^2$. Hence, according to \eqref{eq:GGT}, we have:
		\begin{equation}\label{eq:lambda'-mu}
		  \Psi_{\Lambda',\lambda'}(x')=\Psi_{\Lambda',\mu}(x')\;
		  \e^{\frac{\i}{\hbar}\frac{m}{2t}x'^2}.
		\end{equation}
\end{enumerate}
As a superposition of the three subsequent transformations:
\eqref{eq:mu'-lambda}, \eqref{eq:GFT_free} and
\eqref{eq:lambda'-mu}, we finally obtain:
\begin{align}\label{eq:free_ker}
  \Psi_{\Lambda',\lambda'}(x') &= \sqrt{\frac{m}{\i\hbar t}}
  \int \Psi_{\Lambda,\lambda}(x)\;
    \e^{\frac{\i}{\hbar}\frac{m}{2t}(x-x')^2}\, \sqrt{\d x}\sqrt{\d x'} \ .
\end{align}
To translate this formula to the standard textbook language, where
the quantum state is usually represented by a {\em scalar} wave
function, we make the following {\em Ansatz}:
\begin{equation}\label{ANSATZ}
   \Psi_{\Lambda,\lambda}(x) := \psi_{\Lambda,\lambda}(x)
   \sqrt{\d x} \ ; \qquad
   \Psi_{\Lambda',\lambda'}(x') :=
   \psi_{\Lambda',\lambda'}(x') \sqrt{\d x'} \ .
\end{equation}
As a result, we obtain {\em exactly} the well-known resolution
kernel for the free Schr\"odinger equation:
\begin{align}\label{eq:free_ker-scalar}
  \psi_{\Lambda',\lambda'}(x') &= \sqrt{\frac{m}{\i\hbar t}}
  \int \psi_{\Lambda,\lambda}(x)\;
    \e^{\frac{\i}{\hbar}\frac{m}{2t}(x-x')^2}\, {\d x} \ .
\end{align}
This is where the (arbitrary!) choice of a measure on the
configuration space arises. We stress, however, that formula
\eqref{eq:free_ker}, where the quantum state is correctly
represented by a {\em half-density}, is perfectly invariant with
respect to any change of such a measure.

\subsection{Initial value problem for the harmonic oscillator}
\label{sec:InitialProblemForHarmonicOscillator}

The same three steps lead to the resolution kernel for the
harmonic oscillator. We stress, that the information about the
quantum evolution is entirely encoded in the classical evolution
\eqref{eq:osc_solve}. Indeed, the first step consists in the
generalized Galilei transformation:
\[
  (\lambda- \mu')(x)= m\omega \cot\omega t\, x \d x =
  \d\Big(\frac 1 2 m\omega \cot\omega t\, x^2\Big) \ ,
\]
and, whence, according to  \eqref{eq:GGT} we have:
\begin{equation}\label{eq:mu'-lambda_osc}
  \Psi_{\Lambda,\mu'}(x)=\Psi_{\Lambda,\lambda}(x)\;
  \e^{\frac{\i}{\hbar} \frac 1 2 m\omega \cot\omega t\, x^2} \ .
\end{equation}
Then, we apply the generalized Fourier transformation
\eqref{eq:GFT}:
\begin{align}\label{eq:GFT_osc}
  \Psi_{\Lambda',\mu}(x') &= \int \Psi_{\Lambda,\mu'}(x)\;
    \e^{-\frac{\i}{\hbar}\frac{m\omega}{\sin\omega t}x'x}
    \sqrt{\frac{m\omega}{\i\hbar\sin\omega t}}\,\sqrt{\d x}\sqrt{\d x'},
\end{align}
because $\frac{1}{\i\hbar}\Omega=\frac{1}{\i\hbar} \d p\land \d x
= \frac{m\omega}{\i\hbar\sin\omega t} \d x'\land \d x$. Finally,
we recalculate the wave function from the reference $\mu'$ to the
reference $\lambda$. Formula \eqref{eq:osc_solve} implies:
\[
  (\mu - \lambda')(x')= m\omega \cot\omega t\,x'\d x' =
  \d\Big(\frac 1 2 m\omega \cot\omega t\,x'^2\Big) \ ,
\]
and, consequently:
\begin{equation}\label{eq:lambda'-mu_osc}
  \Psi_{\Lambda',\lambda'}(x')=\Psi_{\Lambda',\mu}(x')\;
  \e^{\frac{\i}{\hbar}\frac 1 2 m\omega \cot\omega t\,x'^2} \ .
\end{equation}
Superposition of the three transformations:
\eqref{eq:mu'-lambda_osc}, \eqref{eq:GFT_osc} and
\eqref{eq:lambda'-mu_osc} gives us:
\begin{align}\label{eq:oscillator_ker}
  \Psi_{\Lambda',\lambda'}(x')
  &= \sqrt{\frac{m\omega}{\i\hbar\sin\omega t}} \int \Psi_{\Lambda,\lambda}(x)
   \; \e^{\frac{\i}{\hbar}
      \frac 1 2 m\omega \left(\cot\omega t(x^2+x'^2)
      -\frac{2x'x}{\sin\omega t}\right)}\, \sqrt{\d x}\sqrt{\d x'} \ .
\end{align}
Again, when translated to the language of {\em scalar} wave
functions by the {\em Ansatz} \eqref{ANSATZ}, the formula gives us the
resolution kernel for the harmonic oscillator. As already
discussed in Section \ref{sec:FFThod}, it is equal to the
fractional Fourier transform \eqref{eq:FrFT}.

\subsection{Initial problem for the charged particle in
constant electric field}
\label{sec:InitialProblemForParticleInConstantElectricField}

Analogously, we consider the case of the charged particle in
constant electric field. Again, using only the classical evolution
\eqref{eq:el_solve} we recover its quantum version in the
following three steps: 1) the Galilei transformation:
\[
  (\lambda- \mu')(x)= \Big(\frac{m}{t}x +\frac{eEt}{2}\Big)\d x =
  \d\Big(\frac{m}{2t}x^2+\frac{eEt}{2}x\Big) \ ,
\]
which implies:
\begin{equation}\label{eq:mu'-lambda_ele}
  \Psi_{\Lambda,\mu'}(x)=\Psi_{\Lambda,\lambda}(x)\;
  \e^{\frac{\i}{\hbar}(\frac{m}{2t}x^2+\frac{eEt}{2}x)} \ ;
\end{equation}
2) the generalized Fourier transformation (which looks similarly
as in case of the free particle, cf. \eqref{eq:GFT_free})
\begin{align}\label{eq:GFT_ele}
  \Psi_{\Lambda',\mu}(x') &= \int \Psi_{\Lambda,\mu'}(x)\;
  \e^{-\frac{\i}{\hbar}\frac{m}{t}x'x}
    \sqrt{\frac{m}{\i\hbar t}}\,\sqrt{\d x}\sqrt{\d x'} \ ,
\end{align}
because $\frac{1}{\i\hbar}\Omega=\frac{1}{\i\hbar} \d p\land \d
x=\frac{m}{\i\hbar t}\d x'\land \d x$ and 3) once more the Galilei
transformation
\[
  (\mu - \lambda')(x')= \Big(\frac{m}{t}x'+\frac{eEt}{2}\Big)\d x'
  = \d\Big(\frac{m}{2t}x'^2+\frac{eEt}{2}x'\Big),
\]
and, consequently:
\begin{equation}\label{eq:lambda'-mu_ele}
  \Psi_{\Lambda',\lambda'}(x')=\Psi_{\Lambda',\mu}(x')\;
  \e^{\frac{\i}{\hbar}(\frac{m}{2t}x'^2+\frac{eEt}{2}x')} \ .
\end{equation}
Superposing the three transformations: \eqref{eq:mu'-lambda_ele},
\eqref{eq:GFT_ele} and \eqref{eq:lambda'-mu_ele}, we get
\begin{align}\label{eq:ele_ker}
  \Psi_{\Lambda',\lambda'}(x')
  &= \sqrt{\frac{m}{\i\hbar t}} \int \Psi_{\Lambda,\lambda}(x)\;
    \e^{\frac{\i}{\hbar}\frac{m}{2t}(x-x')^2}
    \e^{\frac{\i}{\hbar}\frac{eEt}{2}(x+x')}\, \sqrt{\d x}\sqrt{\d x'} \ .
\end{align}
The above wave function satisfies the Schr\"odinger equation in
the linear, time-dependent potential
\begin{equation}
      U(x):= -eEx - \frac{e^2 E^2 t^2}{8m} \ .
\end{equation}
As already discussed in Section
\ref{sec:ParticleInConstantElectricFieldVsFreeMotion}, the time
dependence of the potential {\em via} an irrelevant constant
$-\frac{e^2 E^2 t^2}{8m}$ can be removed by an appropriate gauge
transformation. For this purpose the (physically irrelevant) phase
factor
\[
  \exp\Big(-\frac{\i}{\hbar}\frac{e^2 E^2 t^3}{24m}\Big) \ ,
\]
can be applied. Finally, we obtain the formula
\begin{align}\label{eq:ele_ker_correct}
  \Psi_{\Lambda',\lambda'}(x')
  &= \sqrt{\frac{m}{\i\hbar t}} \int \Psi_{\Lambda,\lambda}(x)\;
    \e^{\frac{\i}{\hbar}\frac{m}{2t}(x-x')^2}
    \e^{\frac{\i}{\hbar}\frac{eEt}{2}(x+x')}
    \e^{-\frac{\i}{\hbar}\frac{e^2 E^2 t^3}{24m}}\, \sqrt{\d x}\sqrt{\d x'} \ ,
\end{align}
which, with the help of the {\em Ansatz} \eqref{ANSATZ}, may be
easily translated to the language of scalar wave functions.

\subsection{Initial problem for the charged particle in constant
magnetic field}
\label{sec:InitialProblemForParticleInConstantMagneticField}

For the sake of completeness we discuss also the charged particle
in constant magnetic field. Using classical dynamics
\eqref{eq:mag_solve_xy} and \eqref{eq:mag_solve_pxpy} we obtain:
1) the Galilei transformation from $\lambda=\{p_x=0, p_y=0\}$ to
$\mu'=\{x'=0, y'=0\}$:
\begin{align*}
  (\lambda -\mu')(x,y) &= \frac 1 2 m\omega\cot\frac{\omega t}{2}\, x \d x
  + \frac 1 2 m\omega\cot\frac{\omega t}{2}\, y\d y
  \\
  &= \d\left(\frac 1 4 m\omega\cot\frac{\omega t}{2}\, (x^2+y^2)\right)\;,
\end{align*}
hence
\begin{equation}\label{eq:mu'-lambda_mag}
  \Psi_{\Lambda,\mu'}(x,y)=\Psi_{\Lambda,\lambda}(x,y)\;
  \e^{\frac{\i}{\hbar}\frac 1 4 m\omega\cot\frac{\omega t}{2}\, (x^2+y^2)}\ ;
\end{equation}
2) the generalized Fourier transformation:
\begin{multline}\label{eq:GFT_mag}
  \Psi_{\Lambda',\mu}(x',y')
  = \int \Psi_{\Lambda,\mu'}(x,y)\;
  \e^{-\frac{\i}{\hbar}
    \frac{1}{2}m\omega\big(\cot\frac{\omega t}{2}\, x'x - y'x + x'y
    +\cot\frac{\omega t}{2}\, y'y\big)}
  \\
  \times\sqrt{\frac{m^2\omega^2}{2\big(\i\hbar\sin\frac{\omega t}{2}\big)^2}}\,
   \sqrt{\d x \d y}\sqrt{\d x' \d y'}\;,
\end{multline}
because $\frac{1}{\i\hbar}\Omega=\frac{1}{\i\hbar}\d p_x\land \d
x+\frac{1}{\i\hbar} \d p_y\land \d
y=\frac{m\omega}{2\i\hbar}\big(\cot\frac{\omega t}{2}\, \d x'\land \d x -
\d y'\land \d x + \d x'\land \d y + \cot\frac{\omega t}{2}\, \d
y'\land \d y\big)$ and

\noindent 3) the Galilei transformation from $\mu=\{x=0, y=0\}$ to
$\lambda'=\{p_x'=0, p_y'=0\}$:
\begin{align*}
  (\mu - \lambda')(x',y') &= \frac 1 2 m\omega\cot\frac{\omega t}{2}\, x'\d x'
  + \frac 1 2 m\omega\cot\frac{\omega t}{2}\, y'\d y'
  \\
  &= \d\big(\frac 1 4 m\omega\cot\frac{\omega t}{2}\,(x'^2+y'^2)\big)\;,
\end{align*}
hence
\begin{equation}\label{eq:lambda'-mu_mag}
  \Psi_{\Lambda',\lambda'}(x',y')=\Psi_{\Lambda',\mu}(x',y')\;
  \e^{\frac{\i}{\hbar}\frac 1 4 m\omega\cot\frac{\omega t}{2}\, (x'^2+y'^2)}\;.
\end{equation}
Finally, formulae \eqref{eq:mu'-lambda_mag}, \eqref{eq:GFT_mag} and
\eqref{eq:lambda'-mu_mag} imply:
\begin{multline}\label{eq:mag_ker}
  \Psi_{\Lambda',\lambda'}(x',y')
  = \sqrt{\frac{m^2\omega^2}{2\big(\i\hbar\sin\frac{\omega t}{2}\big)^2}} \\
    \times \int \Psi_{\Lambda,\lambda}(x,y)\;
    \e^{\frac{\i}{\hbar}
      \frac{1}{4}m\omega\cot\frac{\omega t}{2}\big((x-x')^2+(y-y')^2\big)}
    \e^{-\frac{\i}{\hbar}\frac 1 2 m\omega(x'y-y'x)}\,
    \sqrt{\d x \d y}\sqrt{\d x' \d y'} \ ,
\end{multline}
which is the correct resolution kernel of the quantum initial
value problem.

\section{Quantum connection}
\label{sec:Quantum-conn}

Given two mutually transversal, compatible Lagrangian foliations
$\Lambda_1$ and $\Lambda_2$ of the phase space ${\cal P}$, it is
always possible to choose linear canonical variables $( x^i,p_i)$
in such a way that the symplectic form reduces to
\eqref{canonical}. This way our construction does not go beyond
the Heisenberg group. However, if we take $\Lambda_3$ compatible
with $\Lambda_2$, this does not imply compatibility of $\Lambda_1$
with $\Lambda_3$. Transforming the quantum state form $\Lambda_1$
first to $\Lambda_2$ and then from $\Lambda_2$ to $\Lambda_3$ in a
way defined in this paper, we finally obtain transformation
between foliation which may be far from being compatible. As an
example take again $\dim \mathcal{ P}=2$ and consider a generalized Galilei
transformation of the form
\[
    \tilde{p} = p + \varphi (x) \ .
\]
Take
\[
   \Lambda_1 = \{ p = \mbox{const.} \} \ ,\ \   \Lambda_2 = \{ x =
  \mbox{const.}\}  \ ,\ \  \Lambda_3
  = \{ \tilde{p} = \mbox{const.}\} \ .
\]
Because both $(x,p)$ and $(x, \tilde{p})$ are canonical variables, the pairs
$(\Lambda_1 ,\Lambda_2)$ and $(\Lambda_2 ,\Lambda_3)$
are mutually compatible. But $(\Lambda_1 ,\Lambda_3)$ are, in general,
non-compatible\footnote{They are compatible if and only if $\varphi$ is
linear.}.

Another example is provided by any non-linear
transformation of positions: $x := f(X)$. We have
\begin{equation}\label{non-linear1}
    \Omega=\d p\wedge\d x =  \d p \wedge f^\prime(X)\d X
    =  \d P \wedge \d X \ ,
\end{equation}
where $f^\prime(X) := \frac{\d f}{\d X}(X)$,
$P(x,p):= p \cdot \left(f^\prime \circ f^{-1}(x)\right)$.
Take:
\[
   \Lambda_1 = \{ p = \mbox{const.}\} \ ,\ \   \Lambda_2 = \{ x =
  \mbox{const.}\} = \{ X =\mbox{const.} \} \ ,\ \  \Lambda_3
  = \{ P = \mbox{const.}\} \ .
\]
Of course, $(\Lambda_1 ,\Lambda_2)$ and $(\Lambda_2 ,\Lambda_3)$
are pairwise compatible. But, in general, $(\Lambda_1 ,\Lambda_3)$
is not and the canonical transformation $(x,p) \mapsto (X,P)$
may be highly non-linear. Using our techniques we are, however,
able to construct uniquely the quantum counterpart of such non-linear
canonical transformations. This construction can be extended to an
arbitrary sequence of foliations:  $(\Lambda_1,\Lambda_2, \dots
,\Lambda_i , \Lambda_{i+1}, \dots , \Lambda_N)$, such that two
subsequent foliations are mutually compatible.

The following questions arise:
\begin{enumerate}
    \item Starting from a foliation $\Lambda_{initial}$, can we reach
    this way {\em any} foliation $\Lambda_{final}$?
    \item Does the result depend upon the path, if $\Lambda_{final}$
    can be reached from $\Lambda_{initial}$ in two different ways?
\end{enumerate}

Let us consider the infinitesimal version of this problem. In the first
example we put
\[
    \tilde{p} = p + \epsilon \varphi (x) \ ,
\]
and consider the resulting Hamiltonian vector field (i.e.~an ``infinitesimal
canonical transformation'')
\[
   Y_H = \varphi(x) \frac {\partial}{\partial p} \ ,
\]
generated be the ``Hamiltonian function'' of the form: $H(x,p)= h(x)$ such
that $\varphi = - h^\prime$.

In the second example we put
\[
    x = f(X) = X - \epsilon g(X) \ .
\]
Infinitesimal version of the canonical transformation
\[
   (X,P)= \left( f^{-1} (x) , p \cdot \left(f^\prime \circ f^{-1}(x)\right)
   \right) \ ,
\]
can be easily obtained if we observe that (in first order in $\epsilon$) we
have:
\[
   X = f^{-1}(x) \simeq x + \epsilon g(x)
\]
and, consequently,
\[
   P = p \cdot \left(f^\prime \circ f^{-1}(x)\right)  \simeq p \left( 1 -
   \epsilon g^\prime (x)\right) \ .
\]
This corresponds to the vector field
\[
   Y_H = - p \cdot g^\prime (x) \frac {\partial}{\partial p}
   \ + \ g (x)  \frac {\partial}{\partial x}\ ,
\]
generated by the Hamiltonian function of the form:
$H(x,p)= p \cdot g (x)$.

This way we cover infinitesimal transformations generated by Hamiltonian
functions of the type $h(x)$ and $p \cdot g(x)$, where $h$ and $g$ are
arbitrary (non-linear) functions. But, we have
proved in this paper, that {\em linear} symplectic group has an {\em exact}
projective representation in the space of quantum states. Superposing this
representation with the above two types of generators, we conclude that also
generators of the type $h(\xi)$ and $\eta \cdot g (\xi)$, where
$\xi$ and $\eta$ are arbitrary linear combinations of $x$ and $p$, can be
reached this way. It is obvious, that any Hamiltonian can be approximated by
sums of such functions. We see that our construction enables us to lift any
{\em infinitesimal} canonical transformation to the space of quantum states.

The bundle of all possible quantum states over all possible Lagrangian
foliations acquires, therefore, a unitary connection. Consequently, any
one-parameter family of foliations generated from any $\Lambda_{initial}$ by a
classical dynamics in ${\cal P}$) can be ``quantized'', i.e.~lifted to the
space of quantum states.

It can be checked that this connection is non-flat, i.e.~such a quantization
is path-dependent. This means that finite canonical transformations cannot be
quantized, i.e.~the representation of the {\em linear} symplectic group cannot
be extended to a representation of the complete (non-linear) group.

\section{Conclusions}
\label{sec:Conclusions}

In this paper we have proved that for linear systems, the quantum
evolution is uniquely and unambiguously generated by its classical
counterpart in terms of the (appropriately geometrized) Fourier
and Galilei transformations. This construction leads to a unique {\em
projective} representation of the linear symplectic group. This construction
is virtually unknown, although both its ingredients (the standard Fourier
transformation and the generalized Galilei transformation) belong to the
classical repertoire of quantum mechanics. It implies that solutions of any
Schr\"odinger equation corresponding to a {\em linear} classical evolution can
be obtained from solutions of the {\em free} Schr\"odinger equation {\em via}
a local (in space and time) transformation. Such an observation may provide a
valuable mathematical tool in quantum optics.

In case of a generic, non-linear evolution, the above construction cannot work
because the evolution does not preserve the {\em compatibility}
of the corresponding phase-space foliations. Nevertheless, generalized Galilei
transformations allow us to go beyond linear symplectic structure, at least
infinitesimally. This way a (non-flat) connection in the bundle of quantum
states is uniquely constructed. It allows us to ``quantize'' any classical
evolution, i.e.~one-parameter family of symplectomorphisms. The non-flatness
of the connection implies the non-existence of an extension of the above
representation of the linear symplectic group to representation of the
complete symplectic group.

\subsection*{Acknowledgments}

This work was supported in part by the Polish Ministry of Science and Higher
Education grant No. N N201 372736.

\end{document}